\documentclass[superscriptaddress,showpacs,amsmath,amssymb,twocolumn,nofootinbib]{revtex4-1}

\usepackage{amsmath, amsthm, amssymb, bm, braket}
\usepackage[dvips]{graphicx}
\usepackage{color}
\usepackage{wrapfig}
\usepackage{fancybox}

\begin{document}
\title{Correlation Energy of Proton-Neutron Subsystem in Valence Orbit}

\author{Tomohiro Oishi}
\email[Electronic address: ]{toishi@pd.infn.it} %{gz2r2dhh@gmail.com}
\author{Lorenzo Fortunato}
\email[Electronic address: ]{fortunat@pd.infn.it} %{gz2r2dhh@gmail.com}
\affiliation{Department of Physics and Astronomy ``Galileo Galilei'', 
University of Padova, and} %via F. Marzolo 8, I-35131, Padova, Italy}
\affiliation{I.N.F.N. Sezione di Padova, \\
via F. Marzolo 8, I-35131, Padova, Italy}

%\author{Lorenzo Fortunato}
%\email[Electronic address: ]{fortunato.lorenzo@pd.infn.it} %{gz2r2dhh@gmail.com}
%\affiliation{Department of Physics and Astronomy ``Galileo Galilei'', 
%University of Padova, and I.N.F.N. Sezione di Padova, 
%via F. Marzolo 8, I-35131, Padova, Italy}

\renewcommand{\figurename}{FIG.}
\renewcommand{\tablename}{TABLE}

\newcommand{\bi}[1]{\ensuremath{\boldsymbol{#1}}}
\newcommand{\unit}[1]{\ensuremath{\mathrm{#1}}}
\newcommand{\oprt}[1]{\ensuremath{\hat{\mathcal{#1}}}}
\newcommand{\abs}[1]{\ensuremath{\left| #1 \right|}}

\def \beq{\begin{equation}}
\def \eeq{\end{equation}}
\def \beqa{\begin{eqnarray}}
\def \eeqa{\end{eqnarray}}
\def \Schr{Schr\"odinger }
\def \twop{2{\it p}}
\def \pn{{\it pn}}

\def \bir{\bi{r}}
\def \ubir{\bar{\bi{r}}}
\def \bip{\bi{p}}
\def \ubip{\bar{\bi{r}}}

\begin{abstract}
Deuteron correlation energy (DCE) of the valence proton-neutron subsystem 
is evaluated by utilizing a simple three-body model. 
We focus on the $^6$Li and $^{18}$F nuclei assuming 
the doubly-closed core and the valence proton and neutron. 
Two interaction models, schematic density-dependent contact (SDDC) 
and Minnesota potentials, 
are utilized to describe the proton-neutron interaction. 
Evaluating DCE, 
we conclude that the proton-neutron binding in $^6$Li can be 
stronger than its counterpart of a deuteron in vacuum. 
On the other hand, in $^{18}$F, the energetic correlation is remarkably 
weak, and does not favor the bound, deuteron-like configuration. 
This significant difference between two systems can be understood from a 
competition between the proton-neutron kinetic and pairing energies, 
which are sensitive to the spatial extension of the wave function. 
This result indicates a remarkable dependence of the deuteron 
correlation to its environment and the valence orbits. 
%Toward the more realistic phenomenology, 
%several improvements of the model are proposed. 
\end{abstract}

\maketitle

%%%%%%%%%%%%%%%%%%%%%%%%%%%%%%%%%%%%%%%%%%%%%%%%%%%%%%%%%%%%%%%%%%%%%%%%%%%
\section{Introduction} \label{Sec:intro}
Deuteron ($J^{\pi}=1^+,S=1$) is 
the only possible bound system of two nucleons in vacuum. 
This common sentence in nuclear physics indicates that, 
in the spin-triplet (isospin-singlet) channel, 
nuclear attraction is stronger than that in the 
spin-singlet (isospin-triplet) channel. 
In spite of this unique importance, the spin-triplet 
proton-neutron (\pn) subsystem in finite nuclei has been 
less investigated \cite{1992Csoto,1981Evans,1998Poves,1999Goodman,2010Bert,2011Bert} 
than the spin-singlet pair of the same 
type of nucleons \cite{05BB,13BZ,03Dean_rev,03Bender_rev}.

Thanks to the recent developments of radioactive isotope-beam experiments, 
the access to the spin-triplet \pn-pairing correlation in $N=Z$ nuclei 
is getting possible. 
In these nuclei, in which the valence proton and neutron occupy the 
same major shell, \pn~correlation is expected to be very relevant. 
Comparing this proton-neutron subsystem with that in vacuum, 
a natural question arises: 
``Does a proton-neutron pair at the surface of the nucleus 
behave like a deuteron ?''

The answer to the previous question is, however, not simple to 
address \cite{1969Shanley,1992Csoto,1994Csoto,1999Liset,2007Tursunov,2010Michel,10Ikeda,2012Tani,2013Tani,2014Tani,2014Enyo,2016Masui,2016Tani}. 
In recent theoretical studies, 
it has been shown that the spin-orbit splitting is a key feature of the deuteron 
correlation in nuclei. 
Utilizing the labels, $j_{\gtrless}\equiv l\pm 1/2$ to indicate 
the spin-orbit partners in the same shell, 
the \pn~correlation becomes enhanced when the energy gap 
between $j_>$ and $j_<$ is small \cite{1998Poves,2013Tani,2014Tani}. 
Indeed, a strong spin-triplet \pn~coupling, 
possibly with spatial localization 
that indicates a sort of deuteron condensation at the surface of the nucleus, 
has been predicted \cite{2014Tani,2014Enyo,2016Masui}. 
In Ref. \cite{2016Masui}, the importance of mixing with continuum states 
in the \pn~correlation was also shown. 
The quasi-deuteron configuration \cite{1999Liset}, 
as well as the isospin-singlet condensate \cite{2010Bert,2011Bert}, 
in heavy nuclei have been discussed with similar intents. 
It is worthwhile to remind that a similar discussion on the 
spin-singlet dineutron and diproton correlation has been also carried 
out \cite{01Oert,2005Matsuo,2006Matsuo,07Bertulani_76,07Marg,08Marg,10Kiku,2005HS,07Hagi_01,07Hagi_03,08Hagi,09Dasso,2010Oishi,11KEnyo,13Shim,2014Lorenzo,2016Lay,2016Singh}. 

In spite of all the accumulated knowledge, 
it is still an open question whether the \pn~pair can be 
considered as bound or not in finite nuclei \cite{1977Wildermuth}. 
Especially, its dependence on the selected orbit(s) or 
on the stability of the whole system has not been clarified as yet. 
This information should be essential also for the phenomenology 
of the Gamow-Teller transition \cite{2002Jacek,12Sagawa_GT}, 
nuclear magnetic mode \cite{1999Liset,2007Tursunov,2014Tani} 
and meta-stable states \cite{2007Tursunov,2010Michel}.

In this article, we present 
a phenomenological evaluation of the so-called 
deuteron-correlation energy (DCE). 
We also investigate 
its sensitivity to the properties of finite nuclei 
by comparing two systems: $^6$Li and $^{18}$F. 

Concerning the first topic, 
we employ core-orbital three-body model, assuming a doubly-closed core plus 
the valence proton and neutron. 
Then, we evaluate the mean energy of the partial \pn~Hamiltonian, 
which can be well separated from the total energy. 
An advantage of our definition of 
DCE is that it becomes equivalent to the deuteron 
binding energy, if the \pn~subsystem is isolated. 
Thus, it gives us direct information on the changes that 
appear in finite systems with respect to the counterpart in vacuum.

For the second topic, 
we discuss the deuteron correlation in 
valence orbits in light $N=Z$ nuclei, $^{6}$Li and $^{18}$F. 
By evaluating DCE in three-body systems, 
we can investigate the sensitivity of deuteron-like subsystem 
to its environment. 
Here we point out a qualitative difference between these two 
systems: in $^{18}$F, the $A=17$ core-nucleon subsystem are bound, 
whereas this is not the case for the $A=5$ subsystems of $^6$Li. 
Thus, it is suitable to compare the deuteron 
correlation in systems where it is strongly or weakly bound. 
We also discuss the reliability of several interaction 
models, which play an essential role in the 
deuteron correlation problem. 
For simplicity, in this article, we utilize only the two-body 
interactions, which should be tuned for each subsystem.

In Sec.\ref{Sec:2}, the formalism of our three-body model is presented. 
Our results and discussion for $^{6}$Li are also summarized there. 
Section \ref{Sec:3} is devoted to $^{18}$F, with 
a comparison to $^{6}$Li. 
Finally, in Sec.\ref{Sec:4}, we summarize the main points of this article, 
as well as the possible improvements for future studies.

%%%%%%%%%%%%%%%%%%%%%%%%%%%%%%%%%%%%%%%%%%%%%%%%%%%%%%%%%%%%%%%%%%%%%%%%%%%
\section{$^6$Li Nucleus} \label{Sec:2}
\subsection{Three-Body Model}
Our investigation starts with the $^6$Li nucleus, 
employing the core-orbital coordinates, 
$\left\{ \bir_p,\bir_n \right\}$, for the three-body system, $\alpha+p+n$. 
The detailed formalism of these coordinates is summarized in Appendix. 
Within this framework, our three-body Hamiltonian 
is given as, 
\beqa
 H_{3b} &=& h_p + h_n + x_{\rm rec} + v_{\rm p-n}(\bir_p, \bir_n), \nonumber \\
 h_i &=& \frac{p_i^2}{2\mu_i} + V_{c-i}(r_i), \nonumber \\
 x_{\rm rec} &=& \frac{\bi{p}_p \cdot \bi{p}_n}{m_c}~~~({\rm recoil~term}), \label{eq:fevag}
\eeqa
where $i=p$ and $n$ for the valence proton and neutron, respectively. 
Here, $\bir_i$ is the relative coordinate between the core and the 
$i$-th nucleon. 
Mass parameters are fixed as follows: 
$\mu_i=m_i m_c/(m_i + m_c)$, 
$m_p=938.272$ MeV$/c^2$, 
$m_n=939.565$ MeV$/c^2$, and 
$m_c = 3727.379$ MeV$/c^2$ ($\alpha$-particle mass). 
Namely, $h_i$ is the single particle (s.p.) 
Hamiltonian between the core and the $i$-th nucleon. 
\begin{figure}[tb] \begin{center}
 \includegraphics[]{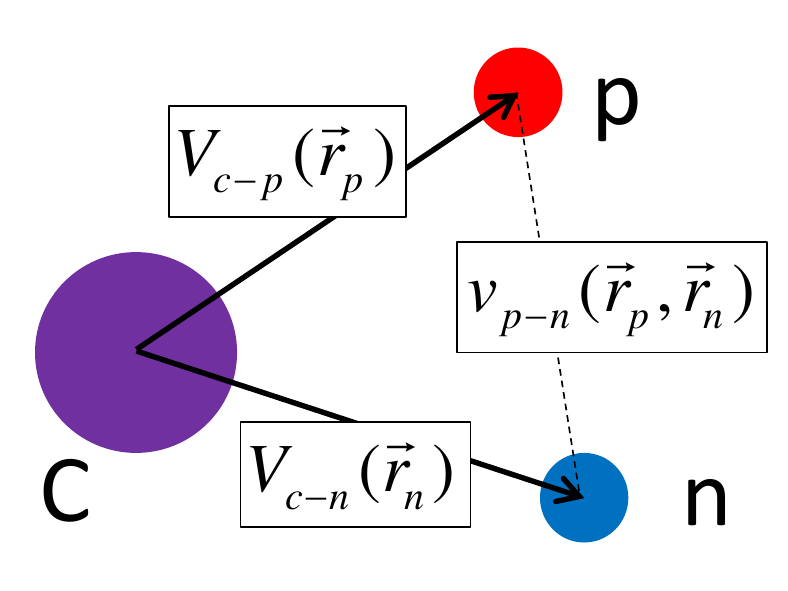}
 \caption{Three-body model with the core and the 
valence proton and neutron. } \label{fig:3}
\end{center} \end{figure}

The core-nucleon potential is taken as 
\beq
 V_{c-i}(\bir_i) = V_{WS}(\bir_i) + V_{Coul}(\bir_i) \delta_{i,p}, \label{eq:mx6}
\eeq
where the Coulomb potential of an uniformly charged sphere with 
radius $R_0$ is included for the core-proton subsystem only. 
For nuclear force, a Woods-Saxon plus spin-orbit potential is employed as 
\beqa
 V_{WS}(r) &=& V_0 f(r) + U_{ls} (\bi{l} \cdot \bi{s}) \frac{1}{r} \frac{df(r)}{dr}, \label{eq:cp-WS} \\
 f(r) &=& \frac{1}{1 + e^{(r-R_0)/a_0}}, \label{eq:coredens}
\eeqa
where $f(r)$ is a standard Fermi profile. 
In this paper, we adopt the parameters as 
$R_0=r_0\cdot 4^{1/3}$, $r_0=1.25$ fm, $a_0=0.65$ fm, $V_0=-47.4$ MeV, and 
$U_{ls}=-0.4092V_0 r_0^2$ \cite{1997EBH,2005HS}. 
From phase-shift analysis, we confirmed that 
this parameter set fairly reproduces the empirical 
$\alpha-n$ and $\alpha-p$ scattering data 
in the $(p_{3/2})$-channel \cite{88Ajzen,02Till,NNDCHP}, 
as summarized in Table \ref{table:fwure}. 

%That is, 
%\beq
% V_{Coul}(r) = \left\{ \begin{array}{cc}
%     \frac{Z_{c}e^2}{4\pi \epsilon_0} \frac{1}{r} & (r > R_0) \\
%     \frac{Z_{c}e^2}{4\pi \epsilon_0} \frac{1}{2R_0} \left(3-\frac{r^2}{R_0^2}\right) & (r \leq R_0)
%    \end{array} \right. ~.
%\eeq

We expand the relevant \pn-states on an uncorrelated basis, 
that is the tensor product of proton and neutron states: 
\beq
 \ket{\Phi^{pn(J,\pi)}_{\kappa \lambda} }
 = \left[ \ket{\phi^p_{\kappa}} \otimes \ket{\phi^n_{\lambda}} \right]^{(J,\pi)}, 
\eeq
where $\kappa$ is the shorthand label for all the quantum numbers 
of the proton states, $\left\{ n_p,l_p,j_p,m_p \right\}$, and 
similarly with $\lambda$ for neutron states. 
Those include the radial quantum number $n$, 
the orbital angular momentum $l$, 
the spin-coupled angular momentum $j$ and 
the magnetic quantum number $m$. 
Each s.p. state satisfies, 
\beqa 
 && h_{p} \phi^{p}_{\kappa }(\bir_{p}) = \epsilon_{\kappa } \phi^{p}_{\kappa }(\bir_{p}), \nonumber \\
 && h_{n} \phi^{n}_{\lambda}(\bir_{n}) = \epsilon_{\lambda} \phi^{n}_{\lambda}(\bir_{n}),
\eeqa 
where $\epsilon_{\kappa(\lambda)}$ is the single-proton (neutron) eigen-energy. 
Notice that these states describe the $A=5$ unbound systems, 
$^5$Li and $^5$He. 
We employ the s.p. states up to the $(h_{11/2,~9/2})$-channel ($l_{\rm max}=5$). %and $j_{\rm max}=2l_{\rm max}+1$. 
We confirmed that this truncation provides a sufficient convergence 
of the ground state energy of $^6$Li. 
In order to take into account the Pauli principle, 
we exclude the first $(s_{1/2})$ state occupied by the core nucleus. 
The continuum s.p. states 
are discretized in the radial box of $R_{\rm box}=20$ fm. 
We also fix the energy cut-off, $E_{\rm cut}=15$ MeV, in this article. 
Because we limit our investigation to the low-energy region only, 
this truncation of model space indeed provides a sufficient convergence 
for our results. 

%Thus, continuum eigenstates of $H_{\rm 3b}$ are also discretized. 
%As we present in Sec.\ref{Subsec:width}, this radial box is sufficiently large 
%to provide a good convergence in terms of the decay width. 

\begin{table}[tb] \begin{center}
\caption{Resonance energy and width obtained with the alpha-nucleon potential. 
Those are evaluated from the scattering 
phase-shift in the $(p_{3/2})$-channel. } \label{table:fwure}
  \catcode`? = \active \def?{\phantom{0}} %define `?' as ' '(one-blank).
  \begingroup \renewcommand{\arraystretch}{1.2}
  \begin{tabular*}{\hsize} { @{\extracolsep{\fill}} cccc } \hline \hline
         &$V_0$  & $E_{\alpha-n},~\Gamma_{\alpha-n}$  & $E_{\alpha-p},~\Gamma_{\alpha-p}$  \\
         &(MeV)  & (MeV)                      & (MeV)                      \\ \hline
  ~This work           &$-47.4$   &$0.77,~0.67$         & $1.61,~1.31$  \\ %~Ref.\cite{14Oishi}  & $1.032,~0.967$         & $1.954,~1.693$  \\
  ~                    &$-49.0$   &$0.54,~0.38$         & $1.37,~0.94$  \\ \hline
  ~Exp.\cite{88Ajzen}  &&$0.89,~0.60$   &$1.97,~(\simeq 1.5)$ \\
  ~Exp.\cite{02Till}   &&$0.798,~0.578$   &$1.69,~1.06$ \\
  ~Exp.\cite{NNDCHP}   &&$0.735,~0.60$   &$1.96(5),~(\simeq 1.5)$~ \\ \hline \hline
  \end{tabular*}
  \endgroup
  \catcode`? = 12 %initialize `?'.
\end{center} \end{table}

\subsection{Proton-Neutron Interactions}
\subsubsection{Schematic Density-Dependent Contact Interaction}
For \pn~subsystem, we employ two simple interaction models in this article. 
Our first choice is the so-called 
schematic density-dependent contact (SDDC) 
potential \cite{2012Tani,2014Tani}. 
That is, 
\beqa
 v_{\rm p-n}(\bir_p,\bir_n) &=&  w \left( \abs{ \frac{\bir_p+\bir_n}{2} } \right) \cdot \delta(\bir_p-\bir_n), \nonumber \\
 w(r) &=& w_0 \left[ 1 - \eta f(r) \right],  \label{eq:w_SDDC}
\eeqa
where $\eta$ is an adjustment parameter. 
We utilize the same density profile, $f(r)$, 
in Eqs.(\ref{eq:cp-WS}) and (\ref{eq:w_SDDC}), 
in order to take the schematic density-dependence into account. 
Notice that $w(r \rightarrow 0) = w_0$ for an isolated 
proton-neutron pair from the core. 
Thus, its bare strength, $w_0$, should be determined 
consistently to the energy cutoff, $E_{\rm cut}$, 
and the vacuum scattering length, $a_v^{(S)}$ \cite{1991BE,1997EBH}: 
\beq
 a_v^{(S)} = \left[ \frac{2k_{\rm cut}}{\pi} 
 + 4\pi \frac{\hbar^2}{2\mu_{\rm p-n} w^{(S)}_0} \right]^{-1}~~({\rm fm}), 
\eeq
or equivalently, 
\beq
 w_0^{(S)} = \frac{\hbar^2}{2\mu_{\rm p-n}} \cdot 
               \frac{4\pi^2 a_v^{(S)}}{\pi-2a_v^{(S)} k_{\rm cut}} 
~~({\rm MeV\cdot fm^3}), 
\eeq
where $\mu_{\rm p-n}=m_p m_n/(m_p+m_n)$ and 
$k_{\rm cut}=\sqrt{2\mu_{\rm p-n} E_{\rm cut}}/\hbar$. 
The superscripts $S=0$ and $1$ indicate the spin-singlet and triplet channels, respectively.

In Fig.\ref{fig:3}, the relationship between the bare strength and 
the vacuum scattering length is presented. 
Empirical values are 
$a_v^{(S=0)}=-23.748$ fm and $a_v^{(S=1)}=5.424$ fm for 
the spin-singlet and triplet channels, respectively \cite{1983Dumbrajs,2007Babenko}. 
Since neutron and proton can become bound in the 
spin-triplet channel, 
$a_v^{(1)}$ is positive finite, and 
its corresponding bare strength, 
$w_0\simeq -2600$ MeV$\cdot{\rm fm}^3$ for $E_{\rm cut}=15$ MeV, 
provides a strong attraction in vacuum. 
On the other hand, in the spin-singlet channel, 
$a_v^{(0)}$ stays negative and the pairing attraction 
is incapable to bind the \pn~system. 
\begin{figure}[tb] \begin{center}
 \includegraphics[width = \hsize]{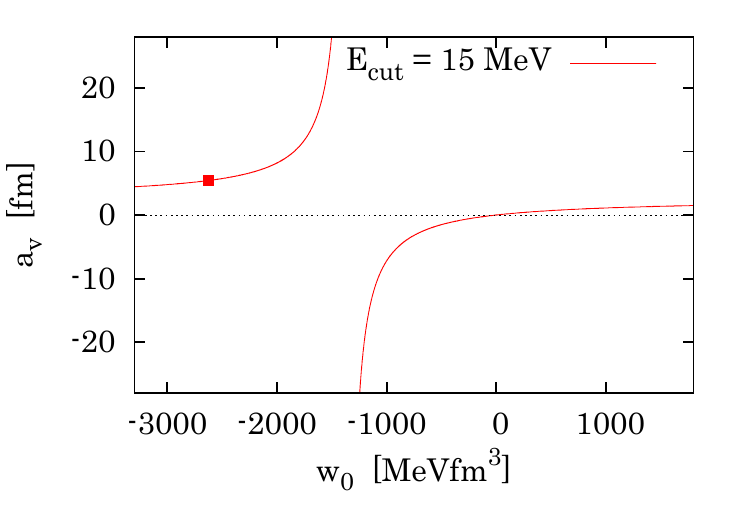} %{fig_01.eps}
 \caption{Relationship between the bare contact strength and 
the vacuum scattering length. 
The empirical value of the spin-triplet \pn-scattering 
length, $a_v^{(S=1)}=5.424$ fm, is indicated with a symbol.} \label{fig:3}
\end{center} \end{figure}

In this article, we only deal with the $J^{\pi}=1^+$ configuration. 
Thus, from angular momentum algebra, the spin-singlet component 
of the \pn-interaction can be neglected \cite{2012Tani}. 

\subsubsection{Minnesota Interaction}
As our second option, 
we employ the spin-triplet Minnesota potential 
for the \pn~subsystem \cite{77Thom,04Suzu,07Hagi_03,10Myo,14Myo_Rev}. 
Using the spin-triplet projection, $\hat{P}_{S=1}$, that is, 
\beqa
 v_{\rm p-n}(\bir_p,\bir_n) &=& V^{(S=1)}_{\rm Min}(\abs{\bir_p-\bir_n}) \hat{P}_{S=1}, \nonumber \\
 V^{(S=1)}_{\rm Min}(r) &=& V_r e^{-K_r r^2} + V_t e^{-K_t r^2}, 
\eeqa
where $V_r=200$ MeV, $V_t=-178$ MeV, $K_r=1.487$ fm$^{-2}$ and 
$K_t=0.639$ fm$^{-2}$ \cite{77Thom}. 
This potential correctly reproduces the deuteron binding energy, 
$\simeq 2.2$ MeV, for the isolated \pn~system.

%\beqa
% S_{\rm pn}(^{6}{\rm Li}) 
% &=& S_{\rm p}(^{6}{\rm Li}) + S_{\rm n}(^{5}{\rm He}) \nonumber \\
% &=& (4433(20)) + (-735(20)) \nonumber \\
% &\simeq & 3698~[{\rm keV}], 
%\eeqa

\subsection{Ground State of $^6$Li}
We solve the ground state (g.s.) of $^6$Li by diagonalizing 
the three-body Hamiltonian. 
This leads to the solution, 
\beq
 \Psi^{(1,+)}_{g.s.} (\bir_p,\bir_n) 
 = \sum_M U_{M} \Phi^{pn(1,+)}_M (\bir_p,\bir_n), 
\eeq
with expansion coefficients, $\{U_M\}$. 
Here $M=\{\kappa,\lambda\}$ is the simplified label for 
the uncorrelated basis. 
In our computation, $M_{\rm max}\simeq 200$ basis states are employed. 

\begin{table*}[tb] \begin{center}
\caption{Ground state of $^6$Li ($1^+$) obtained 
with several \pn-interaction models. 
The empirical three-body binding energy is $E_{3b}=-S_{\rm pn}(^6{\rm Li})=-3.70$ MeV \cite{NNDCHP}. 
The 4 major configurations are also tabulated: 
$\pi$ and $\nu$ indicate the $\alpha$-proton and 
$\alpha$-neutron orbits, respectively. 
$E_v({\rm d})$ is the two-body binding energy of deuteron in vacuum, 
obtained with the Minnesota potential. 
For other quantities, see text for details. 
} \label{table:Jack}
  \catcode`? = \active \def?{\phantom{0}} %define `?' as ' '(one-blank).
  \begingroup \renewcommand{\arraystretch}{1.2}
  \begin{tabular*}{\hsize} { @{\extracolsep{\fill}} ccccc c} \hline \hline
  ~Label &Li-S  &Li-S2  &Li-MO   &Li-MF  &Li-MO2  ~\\
  ~Type of $v_{\rm p-n}$  & \multicolumn{2}{c}{SDDC with $f(r)$ of $\alpha$}   &  \multicolumn{3}{c}{Minnesota for $S=1$}      ~\\
\hline
  ~Adjustment of $v_{\rm p-n}$  &$\eta=1.27$    &$\eta=1.44$    &$f=1$                  &$f=1.13$  &$f=1$  ~\\
  ~$E_v({\rm d})$ (MeV)   &\multicolumn{2}{c}{($a_v^{(S=1)}=5.424$ fm)}   &$-2.22$  &$-4.07$   &$-2.22$  ~\\
  ~$V_0$ of WS Pot. (MeV)       &$-47.4$   &$-49.0$      &$-47.4$  &$-47.4$  &$-49.0$   ~\\
\hline
  ~$E_{3b}=\Braket{H_{3b}}$ (MeV) &$-3.70$   &$-3.70$      &$-2.69$  &$-3.70$  &$-3.64$  ~\\
  ~$\Braket{v_{\rm p-n}}$ (MeV)   &$-8.96$   &$-8.22$      &$-7.65$  &$-8.91$  &$-8.17$  ~\\
  ~$\Braket{x_{\rm rec}}$ (MeV)   &$-0.44$   &$-0.44$      &$-0.38$  &$-0.37$  &$-0.34$  ~\\
  ~$\Braket{\theta_{\rm pn}}$ (deg) & $83.9$ &$83.8$       &$84.6$   &$85.2$   &$85.9$  ~\\
&&&&& \\
  ~$\pi(p_{3/2})\cdot \nu(p_{3/2})$ (\%) &$54.7$ &$56.4$    &$60.4$   &$58.7$  &$62.5$   ~\\
  ~$\pi(p_{1/2})\cdot \nu(p_{3/2})$ (\%) &$18.8$ &$18.1$    &$16.9$   &$17.8$  &$16.5$   ~\\
  ~$\pi(p_{3/2})\cdot \nu(p_{1/2})$ (\%) &$18.3$ &$17.7$    &$16.3$   &$17.3$  &$15.9$   ~\\
  ~$\pi(s_{1/2})\cdot \nu(s_{1/2})$ (\%) &$~2.9$ &$~1.3$    &$~2.3$   &$~2.2$  &$~1.3$   ~\\
&&&&& \\
  ~DCE $\equiv \Braket{h_{p-n}}$  (MeV) &$-4.34$  &$-3.54$       &$-3.21$  &$-4.35$ &$-3.46$  ~\\
  ~$\Braket{h_{c-pn}}$ (MeV) &$+0.64$  &$-0.16$       &$+0.52$  &$+0.65$ &$-0.18$  ~\\
  ~$\Braket{\pi^2_{p-n}/2\mu_{p-n}}$ (MeV)     &$4.62$   &$4.68$      &$4.44$  &$4.56$  &$4.71$    ~\\
  ~$\Braket{\pi^2_{c-pn}/2\mu_{c-pn}}$ (MeV)   &$4.29$   &$4.84$      &$4.43$  &$4.63$  &$5.07$    ~\\
&&&&& \\
  ~$\sqrt{\Braket{\xi^2_{p-n}}}$ (fm)                 &$5.72$    &$5.55$     &$5.61$  &$5.64$  &$5.46$  ~\\
  ~$\sqrt{\Braket{\xi^2_{c-pn}}}$ (fm)                &$3.46$    &$3.42$     &$3.38$  &$3.28$  &$3.11$  ~\\
\hline \hline
  \end{tabular*}
  \endgroup
  \catcode`? = 12 %initialize `?'.
\end{center} \end{table*}

From experimental data \cite{88Ajzen, NNDCHP}, 
the three-body separation energy is, 
\beq
 S_{\rm pn}(^{6}{\rm Li}) = S_{\rm d}(^{6}{\rm Li}) + S_{\rm n}({\rm d}), \nonumber
\eeq
or equivalently, 
\beq
 \phantom{S_{\rm pn}(^{6}{\rm Li})} \nonumber
 = S_{\rm n}(^{6}{\rm Li}) + S_{\rm p}(^{5}{\rm Li}) \simeq  3.70~~({\rm MeV}). 
\eeq
In order to reproduce this empirical energy, 
we employ $\eta=1.27$ for the SDDC potential in Eq. (\ref{eq:w_SDDC}). 

With the Minnesota potential, on the other hand, its original 
parameters fail to reproduce the empirical energy, with 
a positive deviation of almost $1$ MeV. 
Namely, for the fitting, we need an enhancement of the \pn~attraction. 
Thus, in addition, 
we repeat the same calculation but with the enhancement factor, $f=1.13$. 
That is, 
\beq
 v_{\rm p-n}(\bir_p,\bir_n) = f\cdot V^{(S=1)}_{\rm Min}(\abs{\bir_p-\bir_n}) \hat{P}_{S=1}. 
\eeq
This modification, of course, leads to an inconsistency 
with the deuteron energy in vacuum.

In Table \ref{table:Jack}, our results with the SDDC and Minnesota 
interactions (original and fitted) 
are summarized as ``Li-S'', ``Li-MO'', and ``Li-MF'' sets. 
Generally, they well coincide with each other. 
One can find that both \pn~interactions play a 
major role: 
the mean \pn~interaction energy, $\Braket{v_{p-n}}$, 
shows deeply negative values within the g.s. solutions. 

\textcolor{black}{The mean opening angle of \pn, $\Braket{\theta_{pn}}$, 
is less than $90$ degrees in the three cases. 
This indicates a spatial correlation between two nucleons \cite{2016Masui}. 
Indeed, as shown in Ref. \cite{2005HS}, 
this is a product of the mixture of different parities 
with respect of the core-nucleon subsystems: 
if one employs only the odd or even-$l$ 
states in the uncorrelated basis, the resultant 
mean opening angle should be exactly $90$ degrees, 
lacking the spatial correlation. 
In our present result, 
however, this angular correlation is weak compared with the 
isospin-triplet dineutron or diproton correlation \cite{2005HS,2010Oishi,07Bertulani_76,14Oishi}. 
This is consistent with the fact that the contamination from 
channels other than $(p_{3/2})$ and $(p_{1/2})$ is minor 
in this system. }

Comparing the original and enhanced Minnesota cases, 
indicated by Li-MO and Li-MF in Table \ref{table:Jack}, 
the \pn~interaction potential is more attractive in the latter case. 
This is simply 
due to our fitting manipulation to the empirical binding energy. 
However, the above structural information is qualitatively 
similar, and we conclude its weak sensitivity to the binding energy.

\subsection{Deuteron Correlation Energy}
In order to evaluate the deuteron correlation energy (DCE), 
it is more convenient to work with the T-Jacobi coordinates. 
Namely, we can transform our core-orbital coordinates, 
$\left\{ \bir_p,\bir_n \right\}$, to 
the T-Jacobi ones, $\left\{ \bi{\xi}_{p-n}, \bi{\xi}_{c-pn} \right\}$, 
as seen in Fig. \ref{fig:11}. 
Exact formulas can be found in Appendix. 
In these T-Jacobi coordinates, 
our three-body Hamiltonian is decomposed as, 
\beq
 H_{3b} = h_{p-n} + h_{c-pn}, 
\eeq
with two terms, 
\beqa
 h_{p-n} &=& \frac{\pi_{p-n}^2}{2\mu_{p-n}} + v_{\rm p-n}(|\bi{\xi}_{p-n}|), \nonumber \\
 h_{c-pn} &=& \frac{\pi_{c-pn}^2}{2\mu_{c-pn}} + V_{c-p}(\bi{\xi}_{p-n},\bi{\xi}_{c-pn}) \nonumber \\
          & & + V_{c-n}(\bi{\xi}_{p-n},\bi{\xi}_{c-pn}), 
\eeqa
where $\bi{\pi}_{\rm p-n}$ is the relative momentum between 
the valence proton and neutron. 
Thus, $h_{p-n}$ is exactly the \pn-subsystem Hamiltonian, including 
the SDDC or Minnesota interaction.

By taking the expectation value, $\Braket{h_{p-n}}$, 
we can evaluate the deuteron correlation 
inside the three-body system. 
For the \pn~system in vacuum, 
this expectation value of the ground state 
is, of course, the binding energy of deuteron. 
In the following, we employ $\Braket{h_{p-n}}$ as the 
definition of DCE. 
Note that, in some literature \cite{2013Tani,2016Masui}, 
one can find several other definitions of DCE. %for the same purpose. 
\begin{figure}[tb] \begin{center}
 \includegraphics[]{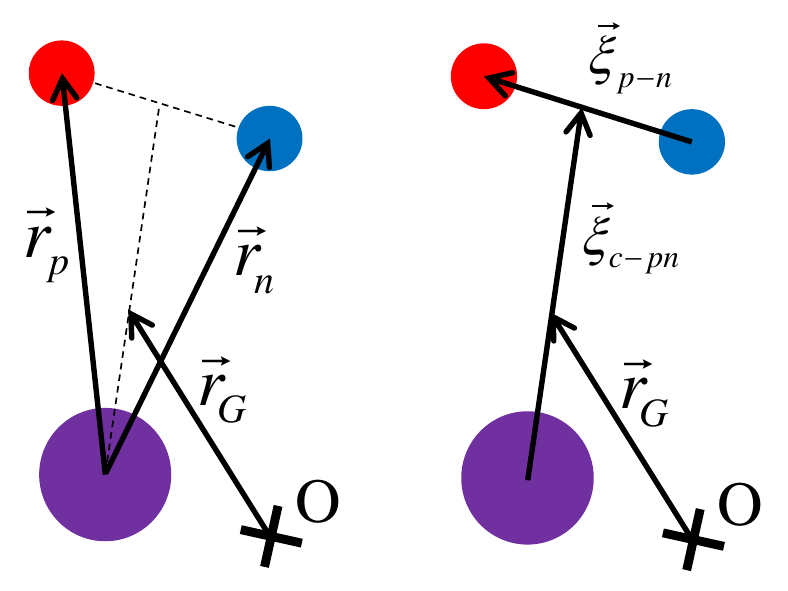}
 \caption{Core-orbital coordinates (left) and 
T-Jacobi coordinates (right). } \label{fig:11}
\end{center} \end{figure}

The result is displayed in the lower half of Table \ref{table:Jack}. 
There, we also tabulate the mean kinetic energies, 
$\Braket{\pi^2_{p-n}/2\mu_{p-n}}$ and $\Braket{\pi^2_{c-pn}/2\mu_{c-pn}}$. 
Notice that 
$\Braket{h_{p-n}}=\Braket{v_{\rm p-n}}+\Braket{\pi^2_{p-n}/2\mu_{p-n}}$ 
from our definition. 
Thus, DCE is the outcome of the competition between 
pairing and kinetic energies, which are 
negative and positive, respectively. 
The dependence of these terms on the selected environment 
is indeed the core of our problem.

From the DCE values, it can be concluded 
that the deuteron subsystem in $^6$Li 
gets an extra binding with respect to its vacuum counterpart, $-2.22$ MeV. 
This is a common feature in both Li-S, Li-MO, and Li-MF cases. 

In the original Minnesota case (Li-MO), we emphasize that 
its parameters are correctly fitted so as to 
reproduce the binding energy, $-2.22$ MeV, if in vacuum. 
Even when using the same parameters, however, 
an enhancement of DCE around the core nucleus is not negligible. 
Thus, our result provides a typical case study, showing 
how the partial two-body system is affected 
by the presence of the third cluster. 

%the change of the partial two-body system affected by the core as another ingredient. 

%It is also worthwhile to compare the SDDC (Li-S) 
%and the enhanced Minnesota (Li-MF) cases. 
%We can conclude that, as long as the \pn~interaction is 
%fitted to the standard binding energy, 
%DCE has almost the same value. 
%This is independent of which interaction model we employ. 

Beside these interesting results, however, we should face 
one shortcoming, 
namely the instability problem in the core-\pn~channel: 
the expectation value, $\Braket{h_{c-pn}}$, is 
positive both in the SDDC and Minnesota cases. 
Thus, the core-\pn~subsystem should be unbound with 
our parameters. 
\textcolor{black}{To remedy this problem, we employ a slightly 
deeper Woods-Saxon potential for the core-nucleon channels: 
$V_0=-49.0$ MeV in Eq.(\ref{eq:cp-WS}), whereas the other parameters 
are not changed. 
With this potential, the core-nucleon levels are 
slightly deviated from the experimental data, but 
the whole picture still keeps a qualitative consistency: 
both core-proton and core-neutron states are broad resonances, 
as seen in Table \ref{table:fwure} ($V_0=-49.0$ MeV). }
%This is qualitatively against the empirical data, where 
%the $\alpha-d$ threshold is located at $1.47$ MeV above 
%the g.s. of $^6$Li. 

\textcolor{black}{In ``Li-S2'' and ``Li-MO2'' sets in Table \ref{table:Jack}, 
our results with the deeper core-nucleon potential 
are summarized. 
In order to reproduce the three-body binding energy there, 
the SDDC interaction needs to be refitted ($\eta=1.44$), 
whereas the Minnesota can be unchanged from its original value. 
Eventually, the core-\pn~channel is stable with negative 
mean energies. 
Furthermore, also in these cases, our previous 
statement can be kept: 
the \pn~subsystem is more strongly bound (about $50$ \%) than in vacuum. 
Consequently, in all the calculations we have performed, 
an enhancement of DCE has been observed. }

\subsection{Geometric Structure}
\textcolor{black}{
In order to evaluate the spatial extent of the wave function, 
we compute the mean relative distances, 
$\Braket{\xi^2_{p-n}}$ and $\Braket{\xi^2_{c-pn}}$. 
From Appendix, the corresponding operators are given by 
\beqa
 && \xi^2_{p-n} = \abs{\bi{\xi}_{p-n}}^2 = \abs{\bir_p-\bir_n}^2, \nonumber \\
 && \xi^2_{c-pn} = \abs{\bi{\xi}_{c-pn}}^2 = \abs{(\bir_p+\bir_n)/2}^2. 
\eeqa
Thus, the mean distances depend on $\Braket{r^2_{p,n}}$ and $\Braket{\bir_p\cdot \bir_n}$. 
Especially for $\Braket{\xi^2_{p-n}}$, 
if the total three-body system is loosely bound with an extended wave function, 
this value is also large, but with a notable exception: 
when the resultant \pn-opening angle is sufficiently narrow, 
then $\Braket{\xi^2_{p-n}}\simeq 0$. 
}

\textcolor{black}{
At the bottom of Table \ref{table:Jack}, our results are tabulated. 
Comparing those with the kinetic energies, 
one can find a common feature: 
when the relative distance gets narrow, its corresponding 
kinetic energy increases. 
This can be naturally understood from the uncertainty principle 
between the relative coordinates and the conjugate momenta. 
}

%---------<Checked with Lorenzo, 6th June, 2017.>

\textcolor{black}{
For the comparison with another system, $^{18}$F, in the next section, 
we point out a general feature of DCE. 
When the total three-body system is loosely bound, 
the \pn-relative distance, $\Braket{\xi^2_{p-n}}$, is large 
if $\cos \Braket{\theta_{\rm pn}}\simeq 0$, 
and consistently, 
the kinetic energy, $\Braket{\pi^2_{p-n}/2\mu_{p-n}}$, becomes small. 
Consequently, the \pn~subsystem can get energetically ``stable'', 
in spite of the loose stability of the whole system. 
In $^6$Li, indeed, 
this kinetic energy is not sufficient to overcome 
the pairing energy, and thus, the \pn~subsystem is quite deeply bound. }

%\textcolor{black}{
%The above common rule is useful for the comparison 
%between different systems. 
%However, in actual calculations limited to the same system, 
%the opening-angle effect and the shift of the 
%pairing energy cannot be negligible, when we 
%change the model parameters. 
%For example, even with the same binding energy 
%in the Li-S and Li-S2 cases, several results 
%are not consistent, due to the different core-\pn~and 
%\pn-pairing potentials. }

%\textcolor{black}{
%It is suggestive to compare the relative distances in $^6$Li with those 
%in the two-neutron Borromean nucleus, $^6$He \cite{2005HS}. 
%In Ref.\cite{2005HS}, the same distances in the $\alpha+n+n$ 
%system are evaluated: $\sqrt{\Braket{\xi^2_{n-n}}} \cong 4.62$ fm and 
%$\sqrt{\Braket{\xi^2_{c-nn}}} \cong 3.63$ fm. 
%In $^6$He, since the opening angle between two neutrons are narrow, 
%$\Braket{\xi^2_{n-n}}$ is smaller than $\Braket{\xi^2_{p-n}}$ in $^6$Li. 
%On the other hand, $\Braket{\xi^2_{c-nn}}$ is larger, consistently 
%to that $^6$He is loosely bound compared with $^6$Li. }

Finally, concerning the more realistic computations, 
of course, we admit that 
further optimization may be considered. 
Those include the exact treatment of the continuum levels in the core-nucleon 
channels \cite{2010Michel,06Aoyama,14Myo_Rev,14Kruppa,2002IdBetan,12Betan,2017IdBetan}, 
as well as the tensor and spin-orbit components in the \pn~interaction \cite{1992Csoto,08Myo}. 
Those are, however, technically demanding and beyond the 
scope of our present model.

\section{$^{18}$F Nucleus} \label{Sec:3}
Next we focus on another system, $^{18}$F, which 
may also support the deuteron correlation around the core nucleus, $^{16}$O. 
A major difference from $^6$Li is that, in the $^{16}$O$-p$ or $^{16}$O$-n$ 
subsystem, there are some bound s.p. orbits. 
Also, the major shell includes $(s_{1/2})$, $(d_{5/2})$ and $(d_{3/2})$. 
Thus, it can be suitable to investigate the sensitivity of 
the deuteron correlation to the valence orbit(s). 

\begin{table}[b] \begin{center}
\caption{Core-nucleon energy levels of $^{17}$O and $^{17}$F with 
respect to the one-nucleon threshold, obtained with 
the core-nucleon potentials in this work. 
The unit is MeV. 
Subscript $r$ indicates the s.p. resonances, and 
$\Gamma$ is the resonance width obtained from 
the scattering phase-shift. 
} \label{table:Peter}
  \catcode`? = \active \def?{\phantom{0}} %define `?' as ' '(one-blank).
  \begingroup \renewcommand{\arraystretch}{1.2}
  \begin{tabular*}{\hsize} { @{\extracolsep{\fill}} ccccc }
  \hline \hline
               &         &\multicolumn{2}{c}{This work}   &Exp.\cite{NNDCHP} ~\\
               &         &default           &$pf$-mix.    &                  ~\\
\hline
  ~$^{16}$O-n   &$\epsilon(2s_{1/2})$  &\multicolumn{2}{c}{$-3.25$}            &$-3.272$  ~\\
               &$\epsilon(1d_{5/2})$  &\multicolumn{2}{c}{$-4.11$}            &$-4.143$  ~\\
               &$\epsilon_r(d_{3/2})$  &\multicolumn{2}{c}{$+0.90$}           &$+0.941$  ~\\
               &                     &\multicolumn{2}{c}{($\Gamma=0.10$)}   &($\Gamma=0.096$)  ~\\
               &$\epsilon_r(f_{7/2})$  &$+7.01$  &$+4.13$   &$-$  ~\\
               &             &($\Gamma=2.74$)   &($\Gamma=0.58$)     &  ~\\
               &$\epsilon_r(f_{5/2})$  &$+11.6$  &$+9.48$  &$-$  ~\\
               &             &($\Gamma=12.1$)   &($\Gamma=6.04$)     &  ~\\
               &$\epsilon_r(p_{3/2})$  &$+2.72$  &$+0.61$           &$-$  ~\\
               &             &($\Gamma=8.60$)   &($\Gamma=1.15$)  &  ~\\
\hline
  ~$^{16}$O-p   &$\epsilon(2s_{1/2})$  &\multicolumn{2}{c}{$-0.13$}     &$-0.105$  ~\\
               &$\epsilon(1d_{5/2})$  &\multicolumn{2}{c}{$-0.55$}     &$-0.600$  ~\\
               &$\epsilon_r(d_{3/2})$  &\multicolumn{2}{c}{$+4.01$}     &$+4.40$  ~\\
               &                      &\multicolumn{2}{c}{($\Gamma=0.89$)}     &($\Gamma=1.53$)  ~\\
               &$\epsilon_r(f_{7/2})$  &$+10.0$  &$+7.22$   &$-$  ~\\
               &             &($\Gamma=4.02$)  &($\Gamma=1.29$)    &  ~\\
               &$\epsilon_r(f_{5/2})$  &$+14.8$  &$+12.6$    &$-$  ~\\
               &             &($\Gamma=14.4$)  &($\Gamma=7.83$)     &  ~\\
               &$\epsilon_r(p_{3/2})$  &no reso-  &$+2.94$           &$-$  ~\\
               &                      &nance     &($\Gamma=3.58$)  &  ~\\
  \hline \hline
  \end{tabular*}
  \endgroup
  \catcode`? = 12 %initialize `?'.
\end{center} \end{table}

\begin{table*}[tb] \begin{center}
\caption{Same to Table \ref{table:Jack} but for the g.s. of $^{18}$F ($1^+$). 
The empirical binding energy is $E_{3b}=-9.75$ MeV \cite{NNDCHP}. %The relative distances, $\xi_{p-n}$ and $\xi_{c-pn}$, are also tabulated. 
} \label{table:Robert}
  \catcode`? = \active \def?{\phantom{0}} %define `?' as ' '(one-blank).
  \begingroup \renewcommand{\arraystretch}{1.2}
  \begin{tabular*}{\hsize} { @{\extracolsep{\fill}} ccccc c}
\hline \hline
  ~Label &F-S  &F-S2  &F-MO   &F-MF  &F-MF2  ~\\
  ~Type of $v_{\rm p-n}$ &\multicolumn{2}{c}{SDDC with $f(r)$ of $^{16}$O}     &\multicolumn{3}{c}{Minnesota for $S=1$}      ~\\
\hline
  ~Adjustment of $v_{\rm p-n}$  &$\eta=1.32$   &$\eta=1.437$                               &$f=1$    &$f=0.67$   &$f=0.59$       ~\\
  ~$E_v({\rm d})$ (MeV) &\multicolumn{2}{c}{($a_v^{(S=1)}=5.424$ fm)} &$-2.22$  &\multicolumn{2}{c}{$>0$ (unbound)}  ~\\
  ~WS Pot.              &default        &$pf$-mix.          &default     &default   &$pf$-mix.  ~\\
\hline
  ~$E_{3b}=\Braket{H_{3b}}$ (MeV)        &$-9.78$  &$-9.74$   &$-13.33$     &$-9.75$  &$-9.72$  ~\\
  ~$\Braket{v_{\rm p-n}}$ (MeV)          &$-7.17$  &$-8.11$   &$-11.32$     &$-6.89$  &$-6.90$  ~\\
  ~$\Braket{x_{\rm rec}}$ (MeV)          &$-0.10$  &$-0.63$   &$-0.09$     &$-0.07$  &$-0.45$  ~\\
  ~$\Braket{\theta_{\rm pn}}$ (deg)      &$87.4$   &$72.3$    &$88.7$       &$88.7$   &$80.0$   ~\\
&&&&& \\
  ~$\pi(d_{5/2})\cdot \nu(d_{5/2})$ (\%)  &$57.9$  &$52.2$    &$48.4$       &$60.7$   &$61.2$   ~\\
  ~$\pi(d_{3/2})\cdot \nu(d_{5/2})$ (\%)  &$13.9$  &$10.9$    &$15.0$       &$10.4$   &$?9.1$  ~\\
  ~$\pi(d_{5/2})\cdot \nu(d_{3/2})$ (\%)  &$13.5$  &$10.2$    &$15.7$       &$10.5$   &$?8.5$  ~\\
  ~$\pi(s_{1/2})\cdot \nu(s_{1/2})$ (\%)  &$10.7$  &$10.8$    &$17.6$       &$16.0$   &$14.6$  ~\\
  ~$\pi(f_{7/2})\cdot \nu(f_{7/2})$ (\%)  &$~0.4$  &$~6.9$    &$~0.2$       &$0.1$    &$?2.8$   ~\\
  ~$\pi(p_{3/2})\cdot \nu(p_{3/2})$ (\%)  &$~0.5$  &$~1.9$    &$~0.3$       &$0.2$    &$?0.7$   ~\\
&&&&& \\
  ~DCE $\equiv \Braket{h_{p-n}}$  (MeV)         &$+0.33$    &$+3.17$     &$-2.05$     &$+1.78$  &$+4.66$  ~\\
  ~$\Braket{h_{c-pn}}$ (MeV)          &$-10.11$  &$-12.91$   &$-11.28$     &$-11.53$ &$-14.38$  ~\\
  ~$\Braket{\pi^2_{p-n}/2\mu_{p-n}}$ (MeV)           &$7.50$  &$11.28$      &$9.27$       &$8.67$   &$11.56$    ~\\
  ~$\Braket{\pi^2_{c-pn}/2\mu_{c-pn}}$ (MeV)          &$6.59$  &$?1.34$      &$8.89$       &$8.49$   &$?4.82$    ~\\
&&&&& \\
  ~$\sqrt{\Braket{\xi^2_{p-n}}}$ (fm)          &$5.09$    &$4.73$     &$4.96$       &$4.95$   &$4.60$  ~\\
  ~$\sqrt{\Braket{\xi^2_{c-pn}}}$ (fm)         &$2.79$    &$3.37$     &$2.58$       &$2.59$   &$2.83$  ~\\
\hline \hline
  \end{tabular*}
  \endgroup
  \catcode`? = 12 %initialize `?'.
\end{center} \end{table*}

\subsection{Model Parameters}
For $^{18}$F, we perform similar calculations but with an 
appropriate change of parameters. 
First, the core-mass parameter, $m_c$, is changed as appropriate. 
In order to take Pauli principle into account, we exclude 
the $(1s_{1/2}),(1p_{3/2}),$ and $(1p_{1/2})$ states, which 
are occupied by the core nucleus. 

\textcolor{black}{
For the core-nucleon interaction, we again adopt the Woods-Saxon potential, 
where Coulomb term is added only for the s.p. proton states. 
In Eq.(\ref{eq:mx6}), some parameters are changed: 
in our default set, 
$R_0=r_0\cdot 16^{1/3}$, 
$V_0^{(l=0)}=-53.1$ MeV, $V_0^{(l\ne0)}=0.99\cdot V_0^{(l=0)}$, and 
$U_{ls}=24.9$ MeV$\cdot$fm$^2$, while $r_0$ and $a_0$ are unchanged. }
In correspondence, the density profile, $f(r)$, in the 
SDDC \pn~interaction is also changed. 

\textcolor{black}{
Additionally to the default set, in Sec. \ref{Sec:Tommy}, 
we also employ $pf$-mixture Woods-Saxon potential. 
There, its depth parameter is modified only for the odd-$l$ channels: 
$V_0^{(l=odd)}=1.188\cdot V_0^{(l=0)}$. }
%Namely, for the odd-$l$ s.p. states, we assume a stronger attraction between the core and nucleons. 

In Table \ref{table:Peter}, 
the core-nucleon levels are summarized. 
Our parameters fairly reproduce the experimental s.p. levels both in 
the proton and neutron channels. 
For resonant channels, 
we also checked the width as obtained from the phase-shift analysis. 
These results approximately coincide with other theoretical 
models \cite{2012Tani,2016Masui}.

\textcolor{black}{Because of the well-determined s.p. levels, 
in contrast to the case of $^6$Li, 
we cannot modify the core-nucleon potentials for the 
major $sd$-shell. 
Thus, in order to reproduce the three-body binding energy, 
the only adoptable way is 
to modify the \pn~interaction parameters. }
For $^{18}$F, this binding energy is measured as $9.75$ MeV \cite{NNDCHP}. 
Thus, the SDDC \pn-pairing interaction is re-adjusted 
with $\eta=1.32$. 
For the Minnesota interaction, on the other hand, we need 
a reduction factor, $f=0.67$, to reproduce this empirical energy 
similarly to Ref.\cite{2016Masui}.

\subsection{Ground State of $^{18}$F}
In ``F-S'', ``F-MO'', and ``F-MF'' sets in Table \ref{table:Robert}, 
our results are summarized in the 
same manner as for $^{6}$Li. 
Generally, \pn~pairing makes a major contribution also in $^{18}$F. 
The mean \pn~interaction energy, $\Braket{v_{\rm p-n}}$, exhausts 
$85$ \% of the three-body binding energy in the 
F-MO case, whereas it amounts to $70$ \% in the other two cases. 

Checking other results in the three cases, the structural properties are 
similar, and not too sensitive to the specific \pn~interaction models. 
There is a small amount of \pn-angular correlation, but not very significant. 
This corresponds to 
a small mixing of the $sd$-shell with other orbits. 
Because of the heavy core, the recoil-term energy is 
almost negligible in this system. 
The mean relative distances also show similar 
values, independently of the \pn~interactions. 
These values are well consistent with the results 
in Ref. \cite{2012Tani}.

\subsection{Energy and Spatial Correlations} \label{Sec:Tommy}
When evaluating the DCE, however, 
the situation becomes contrary to the initial guess of the strong 
deuteron correlation. 
First, in the original Minnesota case (F-MO), DCE is smaller than the value in vacuum. 
Namely, the bound $sd$-shell hardly supports the \pn-energy correlation. 
Furthermore, comparing this DCE with other two cases (F-S and F-MF), where the 
pairing parameters have to be adjusted, 
a drastic change occurs. 
In the F-S and F-MF cases, 
the \pn-subsystem is {\it unstable} around the core, 
because DCE is positive. 
This coincides with the reduction of the pairing attraction 
strength to achieve the empirical binding energy. 
Indeed, the reduced Minnesota force does not support 
the spin-triplet \pn-bound state in vacuum: $E_v({\rm d})>0$. 
Note also that, even with the positive DCE value, 
the whole system can still be stable, as long as 
$\Braket{h_{c-pn}}$ is sufficiently negative.

Until this point, within our simple two-body interaction models, there 
has been no indication of an enhancement of DCE in $^{18}$F, 
showing a remarkable difference from $^6$Li. 
Also, the opening-angle or equivalently the spatial 
correlation is not significant. 
The latter result is in contrast to Ref.\cite{2016Masui}. 
In order to reproduce the spatial \pn~correlation, and to 
investigate its effect on DCE, 
we replace the Woods-Saxon potential with the $pf$-mixture version. 
With this potential, as shown in Table \ref{table:Peter}, 
the odd-$l$ states become closer 
to the Fermi surface, and thus, 
a certain degree of mixing with the g.s. solutions 
is more easily realized. 
%easier to be mixed into the %g.s. solution. 

\textcolor{black}{
In sets ``F-S2'' and ``F-MF2'' of Table \ref{table:Robert}, 
our results with the $pf$-mixture potential are given. 
Indeed, we can find an increase 
of the odd-$l$ contamination. 
Consistently, the opening angle can get closer with this 
potential, as we expected. 
Note also that, for the three-body binding energy, 
\pn-pairing interactions are re-adjusted. 
In these cases with a significant spatial correlation, 
however, the \pn~subsystem is not bound in $^{18}$F. 
Furthermore, its instability becomes enhanced 
compared with the default Woods-Saxon cases (F-S, F-MO and F-MF), as 
indicated by the increase of DCE. }

\textcolor{black}{The instability of the \pn~subsystem 
in the presence of spatial localization 
can be understood from the uncertainty principle. 
When the \pn~subsystem becomes concentrated with the 
narrow distance, $\xi_{p-n}$, the density distribution with respect of 
its conjugate momentum, $\pi_{p-n}$, should be dispersed. 
This leads to the enhancement of the relative kinetic 
energy, $\Braket{\pi^2_{p-n}/2\mu_{p-n}}$, which can be sufficiently large to win the 
\pn-pairing attraction, $\Braket{v_{\rm p-n}}$. 
Consequently, the positively large DCE can be 
attributed to the localized distribution of the probability density. 
Notice also that a good contrast with the wide distribution 
can be found in $^6$Li, where the total system 
is loosely bound compared with $^{18}$F. 
}

%\textcolor{red}{(Better to Remove ?)} 
%Lastly, we make some notes for the spatial localization reported in Ref.\cite{2016Masui}. 
%Indeed, in Ref.\cite{2016Masui}, a core-nucleon potential 
%is more complicated, in order to take into account 
%the exchange effect with the valence nucleons \cite{1991Kaneko}. 
%Evaluation of DCE under this exchange effect, however, 
%requests a technical expansion of our model, and should be beyond the scope of this work. 

\subsection{Complementary Discussions}
Before closing our discussion, 
we present a further comparison of $^6$Li and $^{18}$F nuclei, 
regarding the \pn-correlation dependence on its environment. 
It is also profitable to check the reliability of interaction models, 
as well as its possible improvement. 
\begin{figure}[tb] \begin{center}
 \includegraphics[width=0.95\hsize]{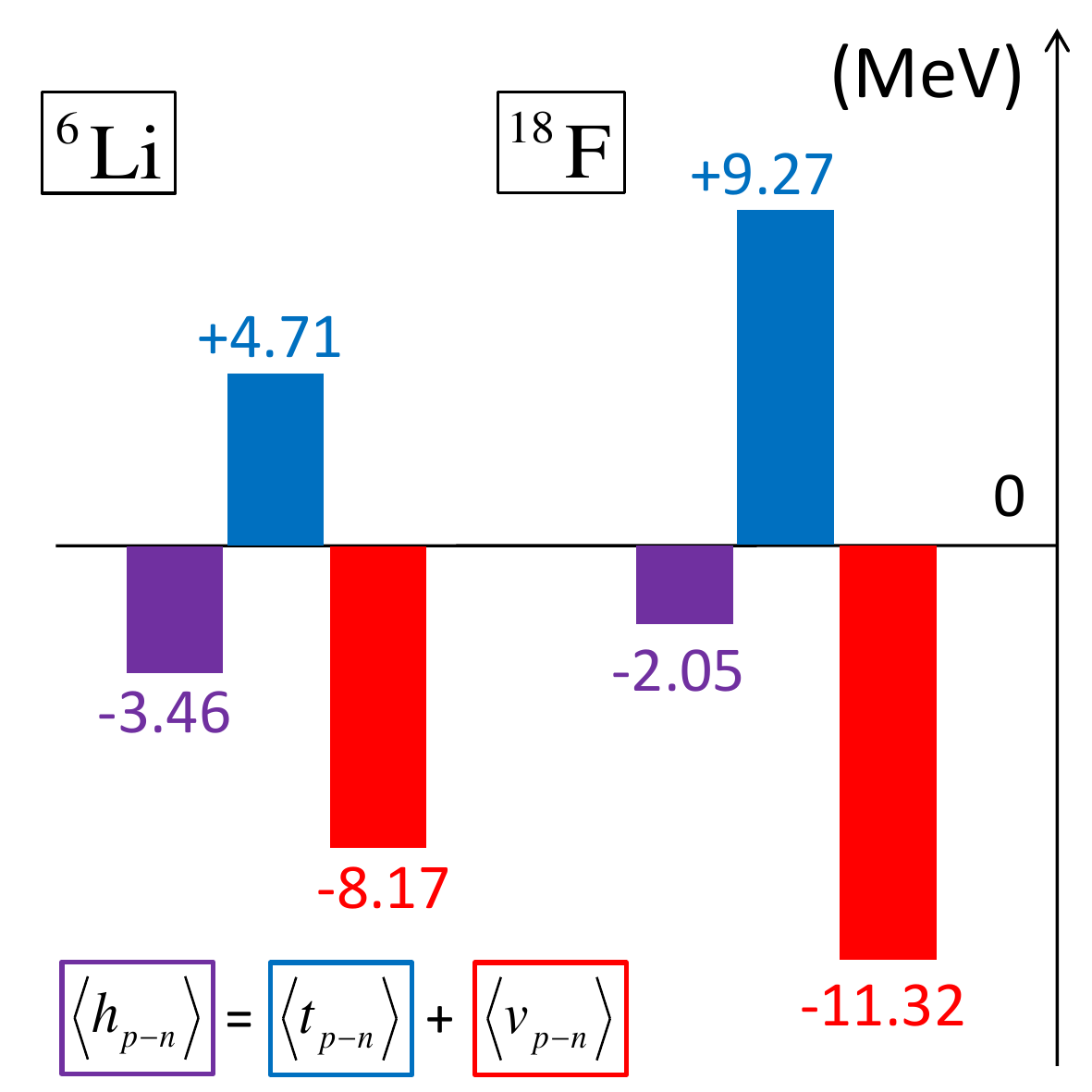}
 \caption{The comparison of DCE in $^6$Li and $^{18}$F, 
where $t_{p-n}\equiv \pi^2_{p-n}/2\mu_{p-n}$. 
These values are taken from ``Li-MO2'' and ``F-MO'' sets 
in Tables \ref{table:Jack} and \ref{table:Robert}, respectively.} \label{fig:DCE}
\end{center} \end{figure}

First, we focus on the original Minnesota cases in two systems, 
Li-MO2 and F-MO, as shown in Fig. \ref{fig:DCE}. 
Here we emphasize that the \pn~interaction operator 
is exactly identical (Minnesota with $f=1$). 
In these cases, the mean pairing energy, 
$\Braket{v_{\rm p-n}}$, clearly depends on the systems. 
This result reflects the effect of the different 
spatial distributions: 
for the short-range attraction like the \pn~interaction, 
its expectation value becomes more negative when the spatial 
distribution is more localized.

\textcolor{black}{In Fig. \ref{fig:DCE}, 
the sensitivity of $\Braket{v_{\rm p-n}}$ to the spatial 
distribution is, however, less drastic than that of the relative 
kinetic energy, $\Braket{\pi^2_{p-n}/2\mu_{p-n}}$. 
This fact causes the stronger DCE of $^6$Li than that of $^{18}$F. 
Notice also that these two terms in $\Braket{h_{p-n}}$ depend on the 
spatial distribution, but in the opposite ways: 
when the mean distance, $\Braket{\xi^2_{p-n}}$, 
gets narrow ($^6$Li $\longrightarrow ^{18}$F), 
$\Braket{v_{\rm p-n}}$ and $\Braket{\pi^2_{p-n}/2\mu_{p-n}}$ 
become negatively and positively enhanced, respectively. }

In sets ``F-MF'' and ``F-MF2'' in Table \ref{table:Robert}, 
in order to reproduce the binding energy of $^{18}$F, 
the \pn-interaction should be reduced, whereas the \pn-kinetic 
operator is common for both $^6$Li and $^{18}$F. 
Consequently, in all the cases we performed, 
the resultant DCE is deeper in the weakly bound 
$p$-shell system, $^6$Li, from the competition between the two energies.

In the Li-MF and F-MF cases, for the Minnesota force fitted 
to the empirical energies of $^6$Li and $^{18}$F, 
the behavior goes in opposite directions: 
$^6$Li requires an enhanced version of 
$V^{(S=1)}_{\rm Min}$, whereas $^{18}$F needs a reduced potential. 
To avoid this case-dependent tuning, %which is less systematically verified, 
it may be necessary to improve 
this \pn~interaction model for future studies. 
%This sophistication may be also profitable to remedy the 
%$\alpha$-\pn~instability problem in $^6$Li, 
%which we already discussed in the last section. 

With the SDDC interaction model, on the other hand, 
we could employ a similar adjusted parameter, $\eta$, 
in both nuclei. 
This advantage comes from the schematic density dependence, 
where the medium effect can be phenomenologically 
taken into account. 
Even with this systematically reliable \pn~interaction, 
consequently, our results show that there is a strong contrast 
between these two nuclei: 
the \pn~subsystem becomes unbound in $^{18}$F, 
whereas it gets deeply bound in $^6$Li.

\section{Summary} \label{Sec:4}
We proposed a direct procedure to evaluate the 
intrinsic deuteron correlation in terms of the subsystem energy. 
By implementing this procedure into a three-body model 
with simple two-body interactions, we discussed the 
\pn~correlation in weakly and strongly bound nuclei. 
From our results, a remarkable sensitivity of DCE to its 
environment is concluded: the \pn~subsystem is more deeply bound in $^6$Li than in 
$^{18}$F. %(SORRY, WRONG !!) therefore the clusterization of the \pn-subsystem into a deuteron-like fragment is less pronounced. 
This can be mainly understood from the uncertainty principle 
between the spatial and momentum distributions: 
because $^{6}$Li is a loosely-bound three-body system, 
its \pn-spatial (momentum) 
distribution can be dispersed (concentrated), and 
thus, the mean \pn-kinetic energy, $\Braket{\pi^2_{p-n}/2\mu_{p-n}}$, gets reduced. 
The comparably small contribution of the \pn-pairing energy, $\Braket{v_{\rm p-n}}$, 
of the SDDC or Minnesota interaction model 
is not sufficient to support a strong \pn~binding in $^{18}$F. 
Our conclusion provides a phenomenological benchmark to discuss 
the \pn~correlation in various situations or/and systems.

There remains several tasks for future studies, 
toward the phenomenological improvement of our model analysis. 
The first possible expansion 
is to implement the spin-orbit or/and tensor forces 
in the \pn~interaction \cite{1992Csoto,08Myo,14Myo_Rev}. 
The sophisticated treatment of continuum 
states may be also profitable for further realistic 
models \cite{14Myo_Rev,89Bohm,05Rotu,06Rotu,06Hagen,06Hagen,2010Michel,14Kruppa,2002IdBetan,12Betan,2017IdBetan}. 
In order to precisely discuss the 
spatial \pn~correlation \cite{2014Enyo,2016Masui}, 
taking the exchange effect of valence particles into account 
might be required \cite{1991Kaneko}. 
With these possible improvements, 
a further evaluation of DCE, covering other systems with 
different spatial extensions, could be reported in future. 
The model-dependence of the \pn-pairing energy also 
needs to be regarded carefully.

Another direction of progress is, as suggested in Ref. \cite{2016Tani}, 
the deuteron emission within a time-dependent 
framework \cite{14Oishi,87Gur,00Talou,09Dasso,12Maru,2015Scamp}. 
From this process, including the \pn-pair tunneling \cite{07Bertulani_34,11Shot}, %\cite{61Bardeen,05Flam}
it has been expected that direct information 
on the \pn~interaction and possibly on correlations 
could be extracted. 
%Because of the high cost of numerical computations for 
%time-dependent processes, 
%it needs to develop a smart multi-body dynamics solver. %\cite{2017FO}. 
%The work toward this direction is in progress now. 

\begin{acknowledgments}
The authors acknowledge the financial support within 
the P.R.A.T. 2015 project {\it IN:Theory} of the University 
of Padova (Project Code: CPDA154713). 
T. Oishi sincerely thanks Yusuke Tanimura, Kouichi Hagino and 
Hiroyuki Sagawa for fruitful discussions. 
The computing facilities offered by CloudVeneto (CSIA Padova and INFN) 
are acknowledged. 
%We also acknowledge the CSC-IT Center for Science Ltd., Finland, 
%for the allocation of computational resources. 
%This work was supported by Academy of Finland and University of 
%Jyv\"{a}skyl\"{a} within the FIDIPRO programme. 
\end{acknowledgments}

\appendix*

\section{Transformation of Coordinates}
In the main text, we employ the core-orbital as well as T-Jacobi 
coordinates for the three-body system. 
In this section, we give a formalism for these transformations. 
First, we need the original coordinates and the conjugate momenta: 
\beq
  \vec{X} \equiv 
  \left[ \begin{array}{c} 
  \bi{x}_p \\ \bi{x}_n \\ \bi{x}_c \end{array} \right], \qquad 
  \vec{Q} \equiv 
  \left[ \begin{array}{c} 
  \bi{q}_p \\ \bi{q}_n \\ \bi{q}_c \end{array} \right]. 
\eeq
In these coordinates, the three-body Hamiltonian is, 
\beq
 H_{3b} = \sum_{i} \frac{\bi{q}_i^2}{m_i}
 + V_{c-p} + V_{c-n} + v_{p-n}, 
\eeq
where $i=p,n$ and $c$ for proton, neutron and core, respectively.

With $3\times 3$ matrix, $U$, the core-orbital 
coordinates can be defined in matrix form: 
\beq
  \vec{R} \equiv 
  \left[ \begin{array}{c} 
  \bir_p \\ \bir_n \\ \bir_G \end{array} \right] 
  = U \vec{X}, \quad
  \vec{P} \equiv 
  \left[ \begin{array}{c} 
  \bip_p \\ \bip_n \\ \bip_G \end{array} \right] 
  = ({}^{t}U)^{-1} \vec{Q}, 
\eeq
where 
\beq
 U = 
 \left( \begin{array}{ccc} 
    1 & 0 & -1 \\
    0 & 1 & -1 \\
    \frac{m_p}{M} & \frac{m_n}{M} & \frac{m_c}{M}
 \end{array} \right), 
\eeq
with $M\equiv \sum_i m_i$ (total mass). 
A schematic view is displayed in Fig.\ref{fig:11}. 
In these coordinates, the Hamiltonian reads 
\beq
 H_{3b} = \frac{p_G^2}{2M} + \frac{p_p^2}{2\mu_p} + \frac{p_n^2}{2\mu_n} 
 + \frac{\bi{p}_p \cdot \bi{p}_n}{m_c} + (potentials), 
\eeq
where the first term represents the center-of-mass kinetic energy, 
that can be neglected. 
This leads to Eq.(\ref{eq:fevag}).

On the other side, T-Jacobi coordinates are given as, 
\beq
 \vec{\Xi} \equiv 
  \left[ \begin{array}{c} 
  \bi{\xi}_{p-n} \\ \bi{\xi}_{c-pn} \\ \bir_G \end{array} \right] 
  = V \vec{X}, \quad
 \vec{\Pi} \equiv 
  \left[ \begin{array}{c} 
  \bi{\pi}_{p-n} \\ \bi{\pi}_{c-pn} \\ \bip_G \end{array} \right] 
  = ({}^{t}V)^{-1} \vec{Q}, 
\eeq
where 
\beq
 V = 
 \left( \begin{array}{ccc} 
    1 & -1 & 0 \\
    \frac{m_p}{m_p+m_n} & \frac{m_n}{m_p+m_n} & -1 \\
    \frac{m_p}{M} & \frac{m_n}{M} & \frac{m_c}{M}
 \end{array} \right). 
\eeq
They are also displayed in Fig.\ref{fig:11}. 
In these T-Jacobi coordinates, the Hamiltonian reads 
\beq
 H_{3b} = \frac{p_G^2}{2M} + \frac{\pi_{p-n}^2}{2\mu_{p-n}} + \frac{\pi_{c-pn}^2}{2\mu_{c-pn}} 
 + (potentials), 
\eeq
with the relative masses, 
\beq
 \frac{1}{\mu_{p-n}} = \frac{m_p + m_n}{m_p m_n}, \quad
 \frac{1}{\mu_{c-pn}} = \frac{1}{m_p+m_n} + \frac{1}{m_c}.
\eeq
Thus, one can separate the \pn-subsystem Hamiltonian, 
$h_{p-n}=\pi_{p-n}^2/2\mu_{p-n}+v_{p-n}$, in these coordinates. 
Notice also that, both in the core-orbital and T-Jacobi coordinate systems, 
the center-of-mass motion is separated.

In order to evaluate the DCE, it is convenient to notice that, 
\beqa
 && \bi{\pi}_{c-pn} = \bip_p + \bip_n, \nonumber \\
 && \bi{\pi}_{p-n} = \frac{m_n\bip_p - m_p\bip_n}{m_p+m_p}, 
\eeqa
as well as, 
\beqa
 \frac{\pi_{c-pn}^2}{2\mu_{c-pn}} &=& 
    \frac{M}{2m_c(m_p+m_n)} (p_p^2 + p_n^2 + 2\bip_p \cdot \bip_n), \nonumber \\
 \frac{\pi_{p-n}^2}{2\mu_{p-n}} &=&
    \frac{m_n p^2_p}{2m_p(m_p+m_n)} + \frac{m_p p^2_n}{2m_n(m_p+m_n)} \nonumber \\
 && -\frac{\bip_p \cdot \bip_n}{m_p+m_n}, 
\eeqa
for the core-orbital and T-Jacobi kinetic operators. 

%The general transformation 
%to $\vec{R}$ and $\vec{P}$ is given as, 
%\beqa
%  && \vec{R} \equiv U \vec{X} \quad \Longleftrightarrow \quad 
%     \vec{X} = U^{-1} \vec{R}, \\
%  && \vec{P} = ({}^{t}U)^{-1} \vec{Q} \quad \Longleftrightarrow \quad 
%     \vec{Q} = {}^{t}U \vec{P}. 
%\eeqa

%\input{bnp_v041.tex}


%---
%\bibliographystyle{apsrev4-1}
%\bibliography{zb_all_18May2017,zb_new}

\begin{thebibliography}{72}%
\makeatletter
\providecommand \@ifxundefined [1]{%
 \@ifx{#1\undefined}
}%
\providecommand \@ifnum [1]{%
 \ifnum #1\expandafter \@firstoftwo
 \else \expandafter \@secondoftwo
 \fi
}%
\providecommand \@ifx [1]{%
 \ifx #1\expandafter \@firstoftwo
 \else \expandafter \@secondoftwo
 \fi
}%
\providecommand \natexlab [1]{#1}%
\providecommand \enquote  [1]{``#1''}%
\providecommand \bibnamefont  [1]{#1}%
\providecommand \bibfnamefont [1]{#1}%
\providecommand \citenamefont [1]{#1}%
\providecommand \href@noop [0]{\@secondoftwo}%
\providecommand \href [0]{\begingroup \@sanitize@url \@href}%
\providecommand \@href[1]{\@@startlink{#1}\@@href}%
\providecommand \@@href[1]{\endgroup#1\@@endlink}%
\providecommand \@sanitize@url [0]{\catcode `\\12\catcode `\$12\catcode
  `\&12\catcode `\#12\catcode `\^12\catcode `\_12\catcode `\%12\relax}%
\providecommand \@@startlink[1]{}%
\providecommand \@@endlink[0]{}%
\providecommand \url  [0]{\begingroup\@sanitize@url \@url }%
\providecommand \@url [1]{\endgroup\@href {#1}{\urlprefix }}%
\providecommand \urlprefix  [0]{URL }%
\providecommand \Eprint [0]{\href }%
\providecommand \doibase [0]{http://dx.doi.org/}%
\providecommand \selectlanguage [0]{\@gobble}%
\providecommand \bibinfo  [0]{\@secondoftwo}%
\providecommand \bibfield  [0]{\@secondoftwo}%
\providecommand \translation [1]{[#1]}%
\providecommand \BibitemOpen [0]{}%
\providecommand \bibitemStop [0]{}%
\providecommand \bibitemNoStop [0]{.\EOS\space}%
\providecommand \EOS [0]{\spacefactor3000\relax}%
\providecommand \BibitemShut  [1]{\csname bibitem#1\endcsname}%
\let\auto@bib@innerbib\@empty
%</preamble>
\bibitem [{\citenamefont {Cs\'ot\'o}\ and\ \citenamefont
  {Lovas}(1992)}]{1992Csoto}%
  \BibitemOpen
  \bibfield  {author} {\bibinfo {author} {\bibfnamefont {A.}~\bibnamefont
  {Cs\'ot\'o}}\ and\ \bibinfo {author} {\bibfnamefont {R.~G.}\ \bibnamefont
  {Lovas}},\ }\href {\doibase 10.1103/PhysRevC.46.576} {\bibfield  {journal}
  {\bibinfo  {journal} {Phys. Rev. C}\ }\textbf {\bibinfo {volume} {46}},\
  \bibinfo {pages} {576} (\bibinfo {year} {1992})}\BibitemShut {NoStop}%
\bibitem [{\citenamefont {Evans}\ \emph {et~al.}(1981)\citenamefont {Evans},
  \citenamefont {Dussel}, \citenamefont {Maqueda},\ and\ \citenamefont
  {Perazzo}}]{1981Evans}%
  \BibitemOpen
  \bibfield  {author} {\bibinfo {author} {\bibfnamefont {J.}~\bibnamefont
  {Evans}}, \bibinfo {author} {\bibfnamefont {G.}~\bibnamefont {Dussel}},
  \bibinfo {author} {\bibfnamefont {E.}~\bibnamefont {Maqueda}}, \ and\
  \bibinfo {author} {\bibfnamefont {R.}~\bibnamefont {Perazzo}},\ }\href
  {\doibase http://dx.doi.org/10.1016/0375-9474(81)90278-5} {\bibfield
  {journal} {\bibinfo  {journal} {Nuclear Physics A}\ }\textbf {\bibinfo
  {volume} {367}},\ \bibinfo {pages} {77 } (\bibinfo {year}
  {1981})}\BibitemShut {NoStop}%
\bibitem [{\citenamefont {Poves}\ and\ \citenamefont
  {Martinez-Pinedo}(1998)}]{1998Poves}%
  \BibitemOpen
  \bibfield  {author} {\bibinfo {author} {\bibfnamefont {A.}~\bibnamefont
  {Poves}}\ and\ \bibinfo {author} {\bibfnamefont {G.}~\bibnamefont
  {Martinez-Pinedo}},\ }\href {\doibase
  https://doi.org/10.1016/S0370-2693(98)00538-3} {\bibfield  {journal}
  {\bibinfo  {journal} {Physics Letters B}\ }\textbf {\bibinfo {volume}
  {430}},\ \bibinfo {pages} {203 } (\bibinfo {year} {1998})}\BibitemShut
  {NoStop}%
\bibitem [{\citenamefont {Goodman}(1999)}]{1999Goodman}%
  \BibitemOpen
  \bibfield  {author} {\bibinfo {author} {\bibfnamefont {A.~L.}\ \bibnamefont
  {Goodman}},\ }\href {\doibase 10.1103/PhysRevC.60.014311} {\bibfield
  {journal} {\bibinfo  {journal} {Phys. Rev. C}\ }\textbf {\bibinfo {volume}
  {60}},\ \bibinfo {pages} {014311} (\bibinfo {year} {1999})}\BibitemShut
  {NoStop}%
\bibitem [{\citenamefont {Bertsch}\ and\ \citenamefont {Luo}(2010)}]{2010Bert}%
  \BibitemOpen
  \bibfield  {author} {\bibinfo {author} {\bibfnamefont {G.~F.}\ \bibnamefont
  {Bertsch}}\ and\ \bibinfo {author} {\bibfnamefont {Y.}~\bibnamefont {Luo}},\
  }\href {\doibase 10.1103/PhysRevC.81.064320} {\bibfield  {journal} {\bibinfo
  {journal} {Phys. Rev. C}\ }\textbf {\bibinfo {volume} {81}},\ \bibinfo
  {pages} {064320} (\bibinfo {year} {2010})}\BibitemShut {NoStop}%
\bibitem [{\citenamefont {Gezerlis}\ \emph {et~al.}(2011)\citenamefont
  {Gezerlis}, \citenamefont {Bertsch},\ and\ \citenamefont {Luo}}]{2011Bert}%
  \BibitemOpen
  \bibfield  {author} {\bibinfo {author} {\bibfnamefont {A.}~\bibnamefont
  {Gezerlis}}, \bibinfo {author} {\bibfnamefont {G.~F.}\ \bibnamefont
  {Bertsch}}, \ and\ \bibinfo {author} {\bibfnamefont {Y.~L.}\ \bibnamefont
  {Luo}},\ }\href {\doibase 10.1103/PhysRevLett.106.252502} {\bibfield
  {journal} {\bibinfo  {journal} {Phys. Rev. Lett.}\ }\textbf {\bibinfo
  {volume} {106}},\ \bibinfo {pages} {252502} (\bibinfo {year}
  {2011})}\BibitemShut {NoStop}%
\bibitem [{\citenamefont {Brink}\ and\ \citenamefont {Broglia}(2005)}]{05BB}%
  \BibitemOpen
  \bibfield  {author} {\bibinfo {author} {\bibfnamefont {D.}~\bibnamefont
  {Brink}}\ and\ \bibinfo {author} {\bibfnamefont {R.}~\bibnamefont
  {Broglia}},\ }\href {http://books.google.co.jp/books?id=K2HAuR-cQ4AC} {\emph
  {\bibinfo {title} {Nuclear Superfluidity: Pairing in Finite Systems}}},\
  Cambridge Monographs on Particle Physics, Nuclear Physics and Cosmology\
  (\bibinfo  {publisher} {Cambridge University Press},\ \bibinfo {address}
  {Cambridge, UK},\ \bibinfo {year} {2005})\BibitemShut {NoStop}%
\bibitem [{\citenamefont {Broglia}\ and\ \citenamefont
  {Zelevinsky}(2013)}]{13BZ}%
  \BibitemOpen
  \bibinfo {editor} {\bibfnamefont {R.~A.}\ \bibnamefont {Broglia}}\ and\
  \bibinfo {editor} {\bibfnamefont {V.}~\bibnamefont {Zelevinsky}},\ eds.,\
  \href {http://www.worldscientific.com/worldscibooks/10.1142/8526#t=aboutBook}
  {\emph {\bibinfo {title} {Fifty Years of Nuclear BCS: Pairing in Finite
  Systems}}}\ (\bibinfo  {publisher} {World Scientific, Singapore},\ \bibinfo
  {year} {2013})\BibitemShut {NoStop}%
\bibitem [{\citenamefont {Dean}\ and\ \citenamefont
  {Hjorth-Jensen}(2003)}]{03Dean_rev}%
  \BibitemOpen
  \bibfield  {author} {\bibinfo {author} {\bibfnamefont {D.~J.}\ \bibnamefont
  {Dean}}\ and\ \bibinfo {author} {\bibfnamefont {M.}~\bibnamefont
  {Hjorth-Jensen}},\ }\href {\doibase 10.1103/RevModPhys.75.607} {\bibfield
  {journal} {\bibinfo  {journal} {Rev. Mod. Phys.}\ }\textbf {\bibinfo {volume}
  {75}},\ \bibinfo {pages} {607} (\bibinfo {year} {2003})}\BibitemShut
  {NoStop}%
\bibitem [{\citenamefont {Bender}\ \emph {et~al.}(2003)\citenamefont {Bender},
  \citenamefont {Heenen},\ and\ \citenamefont {Reinhard}}]{03Bender_rev}%
  \BibitemOpen
  \bibfield  {author} {\bibinfo {author} {\bibfnamefont {M.}~\bibnamefont
  {Bender}}, \bibinfo {author} {\bibfnamefont {P.-H.}\ \bibnamefont {Heenen}},
  \ and\ \bibinfo {author} {\bibfnamefont {P.-G.}\ \bibnamefont {Reinhard}},\
  }\href {\doibase 10.1103/RevModPhys.75.121} {\bibfield  {journal} {\bibinfo
  {journal} {Rev. Mod. Phys.}\ }\textbf {\bibinfo {volume} {75}},\ \bibinfo
  {pages} {121} (\bibinfo {year} {2003})}\BibitemShut {NoStop}%
\bibitem [{\citenamefont {Shanley}(1969)}]{1969Shanley}%
  \BibitemOpen
  \bibfield  {author} {\bibinfo {author} {\bibfnamefont {P.~E.}\ \bibnamefont
  {Shanley}},\ }\href {\doibase 10.1103/PhysRev.187.1328} {\bibfield  {journal}
  {\bibinfo  {journal} {Phys. Rev.}\ }\textbf {\bibinfo {volume} {187}},\
  \bibinfo {pages} {1328} (\bibinfo {year} {1969})}\BibitemShut {NoStop}%
\bibitem [{\citenamefont {Cs\'ot\'o}(1994)}]{1994Csoto}%
  \BibitemOpen
  \bibfield  {author} {\bibinfo {author} {\bibfnamefont {A.}~\bibnamefont
  {Cs\'ot\'o}},\ }\href {\doibase 10.1103/PhysRevC.49.3035} {\bibfield
  {journal} {\bibinfo  {journal} {Phys. Rev. C}\ }\textbf {\bibinfo {volume}
  {49}},\ \bibinfo {pages} {3035} (\bibinfo {year} {1994})}\BibitemShut
  {NoStop}%
\bibitem [{\citenamefont {Lisetskiy}\ \emph {et~al.}(1999)\citenamefont
  {Lisetskiy}, \citenamefont {Jolos}, \citenamefont {Pietralla},\ and\
  \citenamefont {von Brentano}}]{1999Liset}%
  \BibitemOpen
  \bibfield  {author} {\bibinfo {author} {\bibfnamefont {A.~F.}\ \bibnamefont
  {Lisetskiy}}, \bibinfo {author} {\bibfnamefont {R.~V.}\ \bibnamefont
  {Jolos}}, \bibinfo {author} {\bibfnamefont {N.}~\bibnamefont {Pietralla}}, \
  and\ \bibinfo {author} {\bibfnamefont {P.}~\bibnamefont {von Brentano}},\
  }\href {\doibase 10.1103/PhysRevC.60.064310} {\bibfield  {journal} {\bibinfo
  {journal} {Phys. Rev. C}\ }\textbf {\bibinfo {volume} {60}},\ \bibinfo
  {pages} {064310} (\bibinfo {year} {1999})}\BibitemShut {NoStop}%
\bibitem [{\citenamefont {Tursunov}\ \emph {et~al.}(2007)\citenamefont
  {Tursunov}, \citenamefont {Descouvemont},\ and\ \citenamefont
  {Baye}}]{2007Tursunov}%
  \BibitemOpen
  \bibfield  {author} {\bibinfo {author} {\bibfnamefont {E.}~\bibnamefont
  {Tursunov}}, \bibinfo {author} {\bibfnamefont {P.}~\bibnamefont
  {Descouvemont}}, \ and\ \bibinfo {author} {\bibfnamefont {D.}~\bibnamefont
  {Baye}},\ }\href {\doibase http://dx.doi.org/10.1016/j.nuclphysa.2007.06.002}
  {\bibfield  {journal} {\bibinfo  {journal} {Nuclear Physics A}\ }\textbf
  {\bibinfo {volume} {793}},\ \bibinfo {pages} {52} (\bibinfo {year}
  {2007})}\BibitemShut {NoStop}%
\bibitem [{\citenamefont {Michel}\ \emph {et~al.}(2010)\citenamefont {Michel},
  \citenamefont {Nazarewicz},\ and\ \citenamefont
  {P\l{}oszajczak}}]{2010Michel}%
  \BibitemOpen
  \bibfield  {author} {\bibinfo {author} {\bibfnamefont {N.}~\bibnamefont
  {Michel}}, \bibinfo {author} {\bibfnamefont {W.}~\bibnamefont {Nazarewicz}},
  \ and\ \bibinfo {author} {\bibfnamefont {M.}~\bibnamefont {P\l{}oszajczak}},\
  }\href {\doibase 10.1103/PhysRevC.82.044315} {\bibfield  {journal} {\bibinfo
  {journal} {Phys. Rev. C}\ }\textbf {\bibinfo {volume} {82}},\ \bibinfo
  {pages} {044315} (\bibinfo {year} {2010})}\BibitemShut {NoStop}%
\bibitem [{\citenamefont {Ikeda}\ \emph {et~al.}(2010)\citenamefont {Ikeda},
  \citenamefont {Myo}, \citenamefont {Kato},\ and\ \citenamefont
  {Toki}}]{10Ikeda}%
  \BibitemOpen
  \bibfield  {author} {\bibinfo {author} {\bibfnamefont {K.}~\bibnamefont
  {Ikeda}}, \bibinfo {author} {\bibfnamefont {T.}~\bibnamefont {Myo}}, \bibinfo
  {author} {\bibfnamefont {K.}~\bibnamefont {Kato}}, \ and\ \bibinfo {author}
  {\bibfnamefont {H.}~\bibnamefont {Toki}},\ }\href {\doibase
  10.1007/978-3-642-13899-7_5} {\emph {\bibinfo {title} {Clusters in Nuclei:
  Di-Neutron Clustering and Deuteron-like Tensor Correlation in Nuclear
  Structure Focusing on $^{11}$Li}}},\ Lecture Notes in Physics, Vol. 818\
  (\bibinfo  {publisher} {Springer-Verlag},\ \bibinfo {address} {Berlin and
  Heidelberg, Germany},\ \bibinfo {year} {2010})\ pp.\ \bibinfo {pages}
  {165--221}\BibitemShut {NoStop}%
\bibitem [{\citenamefont {Tanimura}\ \emph {et~al.}(2012)\citenamefont
  {Tanimura}, \citenamefont {Hagino},\ and\ \citenamefont {Sagawa}}]{2012Tani}%
  \BibitemOpen
  \bibfield  {author} {\bibinfo {author} {\bibfnamefont {Y.}~\bibnamefont
  {Tanimura}}, \bibinfo {author} {\bibfnamefont {K.}~\bibnamefont {Hagino}}, \
  and\ \bibinfo {author} {\bibfnamefont {H.}~\bibnamefont {Sagawa}},\ }\href
  {\doibase 10.1103/PhysRevC.86.044331} {\bibfield  {journal} {\bibinfo
  {journal} {Phys. Rev. C}\ }\textbf {\bibinfo {volume} {86}},\ \bibinfo
  {pages} {044331} (\bibinfo {year} {2012})}\BibitemShut {NoStop}%
\bibitem [{\citenamefont {Sagawa}\ \emph {et~al.}(2013)\citenamefont {Sagawa},
  \citenamefont {Tanimura},\ and\ \citenamefont {Hagino}}]{2013Tani}%
  \BibitemOpen
  \bibfield  {author} {\bibinfo {author} {\bibfnamefont {H.}~\bibnamefont
  {Sagawa}}, \bibinfo {author} {\bibfnamefont {Y.}~\bibnamefont {Tanimura}}, \
  and\ \bibinfo {author} {\bibfnamefont {K.}~\bibnamefont {Hagino}},\ }\href
  {\doibase 10.1103/PhysRevC.87.034310} {\bibfield  {journal} {\bibinfo
  {journal} {Phys. Rev. C}\ }\textbf {\bibinfo {volume} {87}},\ \bibinfo
  {pages} {034310} (\bibinfo {year} {2013})}\BibitemShut {NoStop}%
\bibitem [{\citenamefont {Tanimura}\ \emph {et~al.}(2014)\citenamefont
  {Tanimura}, \citenamefont {Sagawa},\ and\ \citenamefont {Hagino}}]{2014Tani}%
  \BibitemOpen
  \bibfield  {author} {\bibinfo {author} {\bibfnamefont {Y.}~\bibnamefont
  {Tanimura}}, \bibinfo {author} {\bibfnamefont {H.}~\bibnamefont {Sagawa}}, \
  and\ \bibinfo {author} {\bibfnamefont {K.}~\bibnamefont {Hagino}},\ }\href
  {\doibase 10.1093/ptep/ptu056} {\bibfield  {journal} {\bibinfo  {journal}
  {Progress of Theoretical and Experimental Physics}\ }\textbf {\bibinfo
  {volume} {2014}},\ \bibinfo {pages} {053D02} (\bibinfo {year}
  {2014})}\BibitemShut {NoStop}%
\bibitem [{\citenamefont {Kanada-En'yo}\ and\ \citenamefont
  {Kobayashi}(2014)}]{2014Enyo}%
  \BibitemOpen
  \bibfield  {author} {\bibinfo {author} {\bibfnamefont {Y.}~\bibnamefont
  {Kanada-En'yo}}\ and\ \bibinfo {author} {\bibfnamefont {F.}~\bibnamefont
  {Kobayashi}},\ }\href {\doibase 10.1103/PhysRevC.90.054332} {\bibfield
  {journal} {\bibinfo  {journal} {Phys. Rev. C}\ }\textbf {\bibinfo {volume}
  {90}},\ \bibinfo {pages} {054332} (\bibinfo {year} {2014})}\BibitemShut
  {NoStop}%
\bibitem [{\citenamefont {Masui}\ and\ \citenamefont
  {Kimura}(2016)}]{2016Masui}%
  \BibitemOpen
  \bibfield  {author} {\bibinfo {author} {\bibfnamefont {H.}~\bibnamefont
  {Masui}}\ and\ \bibinfo {author} {\bibfnamefont {M.}~\bibnamefont {Kimura}},\
  }\href {\doibase 10.1093/ptep/ptw041} {\bibfield  {journal} {\bibinfo
  {journal} {Progress of Theoretical and Experimental Physics}\ }\textbf
  {\bibinfo {volume} {2016}},\ \bibinfo {pages} {053D01} (\bibinfo {year}
  {2016})}\BibitemShut {NoStop}%
\bibitem [{\citenamefont {Tanimura}\ and\ \citenamefont
  {Sagawa}(2016)}]{2016Tani}%
  \BibitemOpen
  \bibfield  {author} {\bibinfo {author} {\bibfnamefont {Y.}~\bibnamefont
  {Tanimura}}\ and\ \bibinfo {author} {\bibfnamefont {H.}~\bibnamefont
  {Sagawa}},\ }\href {\doibase 10.1103/PhysRevC.93.064319} {\bibfield
  {journal} {\bibinfo  {journal} {Phys. Rev. C}\ }\textbf {\bibinfo {volume}
  {93}},\ \bibinfo {pages} {064319} (\bibinfo {year} {2016})}\BibitemShut
  {NoStop}%
\bibitem [{\citenamefont {von Oertzen}\ and\ \citenamefont
  {Vitturi}(2001)}]{01Oert}%
  \BibitemOpen
  \bibfield  {author} {\bibinfo {author} {\bibfnamefont {W.}~\bibnamefont {von
  Oertzen}}\ and\ \bibinfo {author} {\bibfnamefont {A.}~\bibnamefont
  {Vitturi}},\ }\href {http://stacks.iop.org/0034-4885/64/i=10/a=202}
  {\bibfield  {journal} {\bibinfo  {journal} {Reports on Progress in Physics}\
  }\textbf {\bibinfo {volume} {64}},\ \bibinfo {pages} {1247} (\bibinfo {year}
  {2001})}\BibitemShut {NoStop}%
\bibitem [{\citenamefont {Matsuo}\ \emph {et~al.}(2005)\citenamefont {Matsuo},
  \citenamefont {Mizuyama},\ and\ \citenamefont {Serizawa}}]{2005Matsuo}%
  \BibitemOpen
  \bibfield  {author} {\bibinfo {author} {\bibfnamefont {M.}~\bibnamefont
  {Matsuo}}, \bibinfo {author} {\bibfnamefont {K.}~\bibnamefont {Mizuyama}}, \
  and\ \bibinfo {author} {\bibfnamefont {Y.}~\bibnamefont {Serizawa}},\ }\href
  {\doibase 10.1103/PhysRevC.71.064326} {\bibfield  {journal} {\bibinfo
  {journal} {Phys. Rev. C}\ }\textbf {\bibinfo {volume} {71}},\ \bibinfo
  {pages} {064326} (\bibinfo {year} {2005})}\BibitemShut {NoStop}%
\bibitem [{\citenamefont {Matsuo}(2006)}]{2006Matsuo}%
  \BibitemOpen
  \bibfield  {author} {\bibinfo {author} {\bibfnamefont {M.}~\bibnamefont
  {Matsuo}},\ }\href {\doibase 10.1103/PhysRevC.73.044309} {\bibfield
  {journal} {\bibinfo  {journal} {Phys. Rev. C}\ }\textbf {\bibinfo {volume}
  {73}},\ \bibinfo {pages} {044309} (\bibinfo {year} {2006})}\BibitemShut
  {NoStop}%
\bibitem [{\citenamefont {Bertulani}\ and\ \citenamefont
  {S.~Hussein}(2007)}]{07Bertulani_76}%
  \BibitemOpen
  \bibfield  {author} {\bibinfo {author} {\bibfnamefont {C.~A.}\ \bibnamefont
  {Bertulani}}\ and\ \bibinfo {author} {\bibfnamefont {M.}~\bibnamefont
  {S.~Hussein}},\ }\href {\doibase 10.1103/PhysRevC.76.051602} {\bibfield
  {journal} {\bibinfo  {journal} {Phys. Rev. C}\ }\textbf {\bibinfo {volume}
  {76}},\ \bibinfo {pages} {051602} (\bibinfo {year} {2007})}\BibitemShut
  {NoStop}%
\bibitem [{\citenamefont {Margueron}\ \emph {et~al.}(2007)\citenamefont
  {Margueron}, \citenamefont {Sagawa},\ and\ \citenamefont {Hagino}}]{07Marg}%
  \BibitemOpen
  \bibfield  {author} {\bibinfo {author} {\bibfnamefont {J.}~\bibnamefont
  {Margueron}}, \bibinfo {author} {\bibfnamefont {H.}~\bibnamefont {Sagawa}}, \
  and\ \bibinfo {author} {\bibfnamefont {K.}~\bibnamefont {Hagino}},\ }\href
  {\doibase 10.1103/PhysRevC.76.064316} {\bibfield  {journal} {\bibinfo
  {journal} {Phys. Rev. C}\ }\textbf {\bibinfo {volume} {76}},\ \bibinfo
  {pages} {064316} (\bibinfo {year} {2007})}\BibitemShut {NoStop}%
\bibitem [{\citenamefont {Margueron}\ \emph {et~al.}(2008)\citenamefont
  {Margueron}, \citenamefont {Sagawa},\ and\ \citenamefont {Hagino}}]{08Marg}%
  \BibitemOpen
  \bibfield  {author} {\bibinfo {author} {\bibfnamefont {J.}~\bibnamefont
  {Margueron}}, \bibinfo {author} {\bibfnamefont {H.}~\bibnamefont {Sagawa}}, \
  and\ \bibinfo {author} {\bibfnamefont {K.}~\bibnamefont {Hagino}},\ }\href
  {\doibase 10.1103/PhysRevC.77.054309} {\bibfield  {journal} {\bibinfo
  {journal} {Phys. Rev. C}\ }\textbf {\bibinfo {volume} {77}},\ \bibinfo
  {pages} {054309} (\bibinfo {year} {2008})}\BibitemShut {NoStop}%
\bibitem [{\citenamefont {Kikuchi}\ \emph {et~al.}(2010)\citenamefont
  {Kikuchi}, \citenamefont {Kat\ifmmode~\bar{o}\else \={o}\fi{}}, \citenamefont
  {Myo}, \citenamefont {Takashina},\ and\ \citenamefont {Ikeda}}]{10Kiku}%
  \BibitemOpen
  \bibfield  {author} {\bibinfo {author} {\bibfnamefont {Y.}~\bibnamefont
  {Kikuchi}}, \bibinfo {author} {\bibfnamefont {K.}~\bibnamefont
  {Kat\ifmmode~\bar{o}\else \={o}\fi{}}}, \bibinfo {author} {\bibfnamefont
  {T.}~\bibnamefont {Myo}}, \bibinfo {author} {\bibfnamefont {M.}~\bibnamefont
  {Takashina}}, \ and\ \bibinfo {author} {\bibfnamefont {K.}~\bibnamefont
  {Ikeda}},\ }\href {\doibase 10.1103/PhysRevC.81.044308} {\bibfield  {journal}
  {\bibinfo  {journal} {Phys. Rev. C}\ }\textbf {\bibinfo {volume} {81}},\
  \bibinfo {pages} {044308} (\bibinfo {year} {2010})}\BibitemShut {NoStop}%
\bibitem [{\citenamefont {Hagino}\ and\ \citenamefont {Sagawa}(2005)}]{2005HS}%
  \BibitemOpen
  \bibfield  {author} {\bibinfo {author} {\bibfnamefont {K.}~\bibnamefont
  {Hagino}}\ and\ \bibinfo {author} {\bibfnamefont {H.}~\bibnamefont
  {Sagawa}},\ }\href {\doibase 10.1103/PhysRevC.72.044321} {\bibfield
  {journal} {\bibinfo  {journal} {Phys. Rev. C}\ }\textbf {\bibinfo {volume}
  {72}},\ \bibinfo {pages} {044321} (\bibinfo {year} {2005})}\BibitemShut
  {NoStop}%
\bibitem [{\citenamefont {Hagino}\ \emph {et~al.}(2007)\citenamefont {Hagino},
  \citenamefont {Sagawa}, \citenamefont {Carbonell},\ and\ \citenamefont
  {Schuck}}]{07Hagi_01}%
  \BibitemOpen
  \bibfield  {author} {\bibinfo {author} {\bibfnamefont {K.}~\bibnamefont
  {Hagino}}, \bibinfo {author} {\bibfnamefont {H.}~\bibnamefont {Sagawa}},
  \bibinfo {author} {\bibfnamefont {J.}~\bibnamefont {Carbonell}}, \ and\
  \bibinfo {author} {\bibfnamefont {P.}~\bibnamefont {Schuck}},\ }\href
  {\doibase 10.1103/PhysRevLett.99.022506} {\bibfield  {journal} {\bibinfo
  {journal} {Phys. Rev. Lett.}\ }\textbf {\bibinfo {volume} {99}},\ \bibinfo
  {pages} {022506} (\bibinfo {year} {2007})}\BibitemShut {NoStop}%
\bibitem [{\citenamefont {Hagino}\ and\ \citenamefont
  {Sagawa}(2007)}]{07Hagi_03}%
  \BibitemOpen
  \bibfield  {author} {\bibinfo {author} {\bibfnamefont {K.}~\bibnamefont
  {Hagino}}\ and\ \bibinfo {author} {\bibfnamefont {H.}~\bibnamefont
  {Sagawa}},\ }\href {\doibase 10.1103/PhysRevC.75.021301} {\bibfield
  {journal} {\bibinfo  {journal} {Phys. Rev. C}\ }\textbf {\bibinfo {volume}
  {75}},\ \bibinfo {pages} {021301} (\bibinfo {year} {2007})}\BibitemShut
  {NoStop}%
\bibitem [{\citenamefont {Hagino}\ \emph {et~al.}(2008)\citenamefont {Hagino},
  \citenamefont {Takahashi},\ and\ \citenamefont {Sagawa}}]{08Hagi}%
  \BibitemOpen
  \bibfield  {author} {\bibinfo {author} {\bibfnamefont {K.}~\bibnamefont
  {Hagino}}, \bibinfo {author} {\bibfnamefont {N.}~\bibnamefont {Takahashi}}, \
  and\ \bibinfo {author} {\bibfnamefont {H.}~\bibnamefont {Sagawa}},\ }\href
  {\doibase 10.1103/PhysRevC.77.054317} {\bibfield  {journal} {\bibinfo
  {journal} {Phys. Rev. C}\ }\textbf {\bibinfo {volume} {77}},\ \bibinfo
  {pages} {054317} (\bibinfo {year} {2008})}\BibitemShut {NoStop}%
\bibitem [{\citenamefont {Dasso}\ and\ \citenamefont
  {Vitturi}(2009)}]{09Dasso}%
  \BibitemOpen
  \bibfield  {author} {\bibinfo {author} {\bibfnamefont {C.~H.}\ \bibnamefont
  {Dasso}}\ and\ \bibinfo {author} {\bibfnamefont {A.}~\bibnamefont
  {Vitturi}},\ }\href {\doibase 10.1103/PhysRevC.79.064620} {\bibfield
  {journal} {\bibinfo  {journal} {Phys. Rev. C}\ }\textbf {\bibinfo {volume}
  {79}},\ \bibinfo {pages} {064620} (\bibinfo {year} {2009})}\BibitemShut
  {NoStop}%
\bibitem [{\citenamefont {Oishi}\ \emph {et~al.}(2010)\citenamefont {Oishi},
  \citenamefont {Hagino},\ and\ \citenamefont {Sagawa}}]{2010Oishi}%
  \BibitemOpen
  \bibfield  {author} {\bibinfo {author} {\bibfnamefont {T.}~\bibnamefont
  {Oishi}}, \bibinfo {author} {\bibfnamefont {K.}~\bibnamefont {Hagino}}, \
  and\ \bibinfo {author} {\bibfnamefont {H.}~\bibnamefont {Sagawa}},\ }\href
  {\doibase 10.1103/PhysRevC.82.024315} {\bibfield  {journal} {\bibinfo
  {journal} {Phys. Rev. C}\ }\textbf {\bibinfo {volume} {82}},\ \bibinfo
  {pages} {024315} (\bibinfo {year} {2010})},\ \bibinfo {note} {with
  erratum.}\BibitemShut {Stop}%
\bibitem [{\citenamefont {Kanada-En'yo}\ \emph {et~al.}(2011)\citenamefont
  {Kanada-En'yo}, \citenamefont {Feldmeier},\ and\ \citenamefont
  {Suhara}}]{11KEnyo}%
  \BibitemOpen
  \bibfield  {author} {\bibinfo {author} {\bibfnamefont {Y.}~\bibnamefont
  {Kanada-En'yo}}, \bibinfo {author} {\bibfnamefont {H.}~\bibnamefont
  {Feldmeier}}, \ and\ \bibinfo {author} {\bibfnamefont {T.}~\bibnamefont
  {Suhara}},\ }\href {\doibase 10.1103/PhysRevC.84.054301} {\bibfield
  {journal} {\bibinfo  {journal} {Phys. Rev. C}\ }\textbf {\bibinfo {volume}
  {84}},\ \bibinfo {pages} {054301} (\bibinfo {year} {2011})}\BibitemShut
  {NoStop}%
\bibitem [{\citenamefont {Shimoyama}\ and\ \citenamefont
  {Matsuo}(2013)}]{13Shim}%
  \BibitemOpen
  \bibfield  {author} {\bibinfo {author} {\bibfnamefont {H.}~\bibnamefont
  {Shimoyama}}\ and\ \bibinfo {author} {\bibfnamefont {M.}~\bibnamefont
  {Matsuo}},\ }\href {\doibase 10.1103/PhysRevC.88.054308} {\bibfield
  {journal} {\bibinfo  {journal} {Phys. Rev. C}\ }\textbf {\bibinfo {volume}
  {88}},\ \bibinfo {pages} {054308} (\bibinfo {year} {2013})}\BibitemShut
  {NoStop}%
\bibitem [{\citenamefont {Fortunato}\ \emph {et~al.}(2014)\citenamefont
  {Fortunato}, \citenamefont {Chatterjee}, \citenamefont {Singh},\ and\
  \citenamefont {Vitturi}}]{2014Lorenzo}%
  \BibitemOpen
  \bibfield  {author} {\bibinfo {author} {\bibfnamefont {L.}~\bibnamefont
  {Fortunato}}, \bibinfo {author} {\bibfnamefont {R.}~\bibnamefont
  {Chatterjee}}, \bibinfo {author} {\bibfnamefont {J.}~\bibnamefont {Singh}}, \
  and\ \bibinfo {author} {\bibfnamefont {A.}~\bibnamefont {Vitturi}},\ }\href
  {\doibase 10.1103/PhysRevC.90.064301} {\bibfield  {journal} {\bibinfo
  {journal} {Phys. Rev. C}\ }\textbf {\bibinfo {volume} {90}},\ \bibinfo
  {pages} {064301} (\bibinfo {year} {2014})}\BibitemShut {NoStop}%
\bibitem [{\citenamefont {Lay}\ \emph {et~al.}(2016)\citenamefont {Lay},
  \citenamefont {Alonso}, \citenamefont {Fortunato},\ and\ \citenamefont
  {Vitturi}}]{2016Lay}%
  \BibitemOpen
  \bibfield  {author} {\bibinfo {author} {\bibfnamefont {J.~A.}\ \bibnamefont
  {Lay}}, \bibinfo {author} {\bibfnamefont {C.~E.}\ \bibnamefont {Alonso}},
  \bibinfo {author} {\bibfnamefont {L.}~\bibnamefont {Fortunato}}, \ and\
  \bibinfo {author} {\bibfnamefont {A.}~\bibnamefont {Vitturi}},\ }\href
  {http://stacks.iop.org/0954-3899/43/i=8/a=085103} {\bibfield  {journal}
  {\bibinfo  {journal} {Journal of Physics G: Nuclear and Particle Physics}\
  }\textbf {\bibinfo {volume} {43}},\ \bibinfo {pages} {085103} (\bibinfo
  {year} {2016})}\BibitemShut {NoStop}%
\bibitem [{\citenamefont {Singh}\ \emph {et~al.}(2016)\citenamefont {Singh},
  \citenamefont {Fortunato}, \citenamefont {Vitturi},\ and\ \citenamefont
  {Chatterjee}}]{2016Singh}%
  \BibitemOpen
  \bibfield  {author} {\bibinfo {author} {\bibfnamefont {J.}~\bibnamefont
  {Singh}}, \bibinfo {author} {\bibfnamefont {L.}~\bibnamefont {Fortunato}},
  \bibinfo {author} {\bibfnamefont {A.}~\bibnamefont {Vitturi}}, \ and\
  \bibinfo {author} {\bibfnamefont {R.}~\bibnamefont {Chatterjee}},\ }\href
  {\doibase 10.1140/epja/i2016-16209-8} {\bibfield  {journal} {\bibinfo
  {journal} {The European Physical Journal A}\ }\textbf {\bibinfo {volume}
  {52}},\ \bibinfo {pages} {209} (\bibinfo {year} {2016})}\BibitemShut
  {NoStop}%
\bibitem [{\citenamefont {Wildermuth}\ and\ \citenamefont
  {Tang}(1977)}]{1977Wildermuth}%
  \BibitemOpen
  \bibfield  {author} {\bibinfo {author} {\bibfnamefont {K.}~\bibnamefont
  {Wildermuth}}\ and\ \bibinfo {author} {\bibfnamefont {Y.}~\bibnamefont
  {Tang}},\ }\href {\doibase 10.1007/978-3-322-85255-7} {\emph {\bibinfo
  {title} {A Unified Theory of the Nucleus}}}\ (\bibinfo  {publisher}
  {Springer, Vieweg+Teubner Verlag},\ \bibinfo {year} {1977})\BibitemShut
  {NoStop}%
\bibitem [{\citenamefont {Bender}\ \emph {et~al.}(2002)\citenamefont {Bender},
  \citenamefont {Dobaczewski}, \citenamefont {Engel},\ and\ \citenamefont
  {Nazarewicz}}]{2002Jacek}%
  \BibitemOpen
  \bibfield  {author} {\bibinfo {author} {\bibfnamefont {M.}~\bibnamefont
  {Bender}}, \bibinfo {author} {\bibfnamefont {J.}~\bibnamefont {Dobaczewski}},
  \bibinfo {author} {\bibfnamefont {J.}~\bibnamefont {Engel}}, \ and\ \bibinfo
  {author} {\bibfnamefont {W.}~\bibnamefont {Nazarewicz}},\ }\href {\doibase
  10.1103/PhysRevC.65.054322} {\bibfield  {journal} {\bibinfo  {journal} {Phys.
  Rev. C}\ }\textbf {\bibinfo {volume} {65}},\ \bibinfo {pages} {054322}
  (\bibinfo {year} {2002})}\BibitemShut {NoStop}%
\bibitem [{\citenamefont {Roca-Maza}\ \emph {et~al.}(2012)\citenamefont
  {Roca-Maza}, \citenamefont {Col\`o},\ and\ \citenamefont
  {Sagawa}}]{12Sagawa_GT}%
  \BibitemOpen
  \bibfield  {author} {\bibinfo {author} {\bibfnamefont {X.}~\bibnamefont
  {Roca-Maza}}, \bibinfo {author} {\bibfnamefont {G.}~\bibnamefont {Col\`o}}, \
  and\ \bibinfo {author} {\bibfnamefont {H.}~\bibnamefont {Sagawa}},\ }\href
  {\doibase 10.1103/PhysRevC.86.031306} {\bibfield  {journal} {\bibinfo
  {journal} {Phys. Rev. C}\ }\textbf {\bibinfo {volume} {86}},\ \bibinfo
  {pages} {031306} (\bibinfo {year} {2012})}\BibitemShut {NoStop}%
\bibitem [{\citenamefont {Esbensen}\ \emph {et~al.}(1997)\citenamefont
  {Esbensen}, \citenamefont {Bertsch},\ and\ \citenamefont
  {Hencken}}]{1997EBH}%
  \BibitemOpen
  \bibfield  {author} {\bibinfo {author} {\bibfnamefont {H.}~\bibnamefont
  {Esbensen}}, \bibinfo {author} {\bibfnamefont {G.~F.}\ \bibnamefont
  {Bertsch}}, \ and\ \bibinfo {author} {\bibfnamefont {K.}~\bibnamefont
  {Hencken}},\ }\href {\doibase 10.1103/PhysRevC.56.3054} {\bibfield  {journal}
  {\bibinfo  {journal} {Phys. Rev. C}\ }\textbf {\bibinfo {volume} {56}},\
  \bibinfo {pages} {3054} (\bibinfo {year} {1997})}\BibitemShut {NoStop}%
\bibitem [{\citenamefont {Ajzenberg-Selove}(1988)}]{88Ajzen}%
  \BibitemOpen
  \bibfield  {author} {\bibinfo {author} {\bibfnamefont {F.}~\bibnamefont
  {Ajzenberg-Selove}},\ }\href {\doibase
  http://dx.doi.org/10.1016/0375-9474(88)90124-8} {\bibfield  {journal}
  {\bibinfo  {journal} {Nuclear Physics A}\ }\textbf {\bibinfo {volume}
  {490}},\ \bibinfo {pages} {1 } (\bibinfo {year} {1988})},\ \bibinfo {note}
  {note: several versions with the same title has been published.}\BibitemShut
  {Stop}%
\bibitem [{\citenamefont {Tilley}\ \emph {et~al.}(2002)\citenamefont {Tilley},
  \citenamefont {Cheves}, \citenamefont {Godwin}, \citenamefont {Hale},
  \citenamefont {Hofmann}, \citenamefont {Kelley}, \citenamefont {Sheu},\ and\
  \citenamefont {Weller}}]{02Till}%
  \BibitemOpen
  \bibfield  {author} {\bibinfo {author} {\bibfnamefont {D.}~\bibnamefont
  {Tilley}}, \bibinfo {author} {\bibfnamefont {C.}~\bibnamefont {Cheves}},
  \bibinfo {author} {\bibfnamefont {J.}~\bibnamefont {Godwin}}, \bibinfo
  {author} {\bibfnamefont {G.}~\bibnamefont {Hale}}, \bibinfo {author}
  {\bibfnamefont {H.}~\bibnamefont {Hofmann}}, \bibinfo {author} {\bibfnamefont
  {J.}~\bibnamefont {Kelley}}, \bibinfo {author} {\bibfnamefont
  {C.}~\bibnamefont {Sheu}}, \ and\ \bibinfo {author} {\bibfnamefont
  {H.}~\bibnamefont {Weller}},\ }\href {\doibase
  http://dx.doi.org/10.1016/S0375-9474(02)00597-3} {\bibfield  {journal}
  {\bibinfo  {journal} {Nuclear Physics A}\ }\textbf {\bibinfo {volume}
  {708}},\ \bibinfo {pages} {3 } (\bibinfo {year} {2002})}\BibitemShut
  {NoStop}%
\bibitem [{NND()}]{NNDCHP}%
  \BibitemOpen
  \href {http://www.nndc.bnl.gov/chart/} {}\bibinfo {note} {Data-base ``Chart
  of Nuclides'', National Nuclear Data Center (NNDC);
  http://www.nndc.bnl.gov/chart/}\BibitemShut {NoStop}%
\bibitem [{\citenamefont {Bertsch}\ and\ \citenamefont
  {Esbensen}(1991)}]{1991BE}%
  \BibitemOpen
  \bibfield  {author} {\bibinfo {author} {\bibfnamefont {G.}~\bibnamefont
  {Bertsch}}\ and\ \bibinfo {author} {\bibfnamefont {H.}~\bibnamefont
  {Esbensen}},\ }\href {\doibase
  http://dx.doi.org/10.1016/0003-4916(91)90033-5} {\bibfield  {journal}
  {\bibinfo  {journal} {Annals of Physics}\ }\textbf {\bibinfo {volume}
  {209}},\ \bibinfo {pages} {327 } (\bibinfo {year} {1991})}\BibitemShut
  {NoStop}%
\bibitem [{\citenamefont {Dumbrajs}\ \emph {et~al.}(1983)\citenamefont
  {Dumbrajs}, \citenamefont {Koch}, \citenamefont {Pilkuhn}, \citenamefont
  {Oades}, \citenamefont {Behrens}, \citenamefont {de~Swart},\ and\
  \citenamefont {Kroll}}]{1983Dumbrajs}%
  \BibitemOpen
  \bibfield  {author} {\bibinfo {author} {\bibfnamefont {O.}~\bibnamefont
  {Dumbrajs}}, \bibinfo {author} {\bibfnamefont {R.}~\bibnamefont {Koch}},
  \bibinfo {author} {\bibfnamefont {H.}~\bibnamefont {Pilkuhn}}, \bibinfo
  {author} {\bibfnamefont {G.}~\bibnamefont {Oades}}, \bibinfo {author}
  {\bibfnamefont {H.}~\bibnamefont {Behrens}}, \bibinfo {author} {\bibfnamefont
  {J.}~\bibnamefont {de~Swart}}, \ and\ \bibinfo {author} {\bibfnamefont
  {P.}~\bibnamefont {Kroll}},\ }\href {\doibase
  http://dx.doi.org/10.1016/0550-3213(83)90288-2} {\bibfield  {journal}
  {\bibinfo  {journal} {Nuclear Physics B}\ }\textbf {\bibinfo {volume}
  {216}},\ \bibinfo {pages} {277} (\bibinfo {year} {1983})}\BibitemShut
  {NoStop}%
\bibitem [{\citenamefont {Babenko}(2007)}]{2007Babenko}%
  \BibitemOpen
  \bibfield  {author} {\bibinfo {author} {\bibfnamefont {N.~M.}\ \bibnamefont
  {Babenko}, \bibfnamefont {V.~A.and~Petrov}},\ }\href {\doibase
  10.1134/S1063778807040072} {\bibfield  {journal} {\bibinfo  {journal}
  {Physics of Atomic Nuclei}\ }\textbf {\bibinfo {volume} {70}},\ \bibinfo
  {pages} {669} (\bibinfo {year} {2007})}\BibitemShut {NoStop}%
\bibitem [{\citenamefont {Thompson}\ \emph {et~al.}(1977)\citenamefont
  {Thompson}, \citenamefont {Lemere},\ and\ \citenamefont {Tang}}]{77Thom}%
  \BibitemOpen
  \bibfield  {author} {\bibinfo {author} {\bibfnamefont {D.}~\bibnamefont
  {Thompson}}, \bibinfo {author} {\bibfnamefont {M.}~\bibnamefont {Lemere}}, \
  and\ \bibinfo {author} {\bibfnamefont {Y.}~\bibnamefont {Tang}},\ }\href
  {\doibase http://dx.doi.org/10.1016/0375-9474(77)90007-0} {\bibfield
  {journal} {\bibinfo  {journal} {Nuclear Physics A}\ }\textbf {\bibinfo
  {volume} {286}},\ \bibinfo {pages} {53 } (\bibinfo {year}
  {1977})}\BibitemShut {NoStop}%
\bibitem [{\citenamefont {Suzuki}\ \emph {et~al.}(2004)\citenamefont {Suzuki},
  \citenamefont {Matsumura},\ and\ \citenamefont {Abu-Ibrahim}}]{04Suzu}%
  \BibitemOpen
  \bibfield  {author} {\bibinfo {author} {\bibfnamefont {Y.}~\bibnamefont
  {Suzuki}}, \bibinfo {author} {\bibfnamefont {H.}~\bibnamefont {Matsumura}}, \
  and\ \bibinfo {author} {\bibfnamefont {B.}~\bibnamefont {Abu-Ibrahim}},\
  }\href {\doibase 10.1103/PhysRevC.70.051302} {\bibfield  {journal} {\bibinfo
  {journal} {Phys. Rev. C}\ }\textbf {\bibinfo {volume} {70}},\ \bibinfo
  {pages} {051302} (\bibinfo {year} {2004})}\BibitemShut {NoStop}%
\bibitem [{\citenamefont {Myo}\ \emph {et~al.}(2010)\citenamefont {Myo},
  \citenamefont {Ando},\ and\ \citenamefont {Kat\ifmmode~\bar{o}\else
  \={o}\fi{}}}]{10Myo}%
  \BibitemOpen
  \bibfield  {author} {\bibinfo {author} {\bibfnamefont {T.}~\bibnamefont
  {Myo}}, \bibinfo {author} {\bibfnamefont {R.}~\bibnamefont {Ando}}, \ and\
  \bibinfo {author} {\bibfnamefont {K.}~\bibnamefont {Kat\ifmmode~\bar{o}\else
  \={o}\fi{}}},\ }\href {\doibase
  http://dx.doi.org/10.1016/j.physletb.2010.06.034} {\bibfield  {journal}
  {\bibinfo  {journal} {Physics Letters B}\ }\textbf {\bibinfo {volume}
  {691}},\ \bibinfo {pages} {150 } (\bibinfo {year} {2010})}\BibitemShut
  {NoStop}%
\bibitem [{\citenamefont {Myo}\ \emph {et~al.}(2014)\citenamefont {Myo},
  \citenamefont {Kikuchi}, \citenamefont {Masui},\ and\ \citenamefont
  {Kato}}]{14Myo_Rev}%
  \BibitemOpen
  \bibfield  {author} {\bibinfo {author} {\bibfnamefont {T.}~\bibnamefont
  {Myo}}, \bibinfo {author} {\bibfnamefont {Y.}~\bibnamefont {Kikuchi}},
  \bibinfo {author} {\bibfnamefont {H.}~\bibnamefont {Masui}}, \ and\ \bibinfo
  {author} {\bibfnamefont {K.}~\bibnamefont {Kato}},\ }\href {\doibase
  https://doi.org/10.1016/j.ppnp.2014.08.001} {\bibfield  {journal} {\bibinfo
  {journal} {Progress in Particle and Nuclear Physics}\ }\textbf {\bibinfo
  {volume} {79}},\ \bibinfo {pages} {1 } (\bibinfo {year} {2014})}\BibitemShut
  {NoStop}%
\bibitem [{\citenamefont {Oishi}\ \emph {et~al.}(2014)\citenamefont {Oishi},
  \citenamefont {Hagino},\ and\ \citenamefont {Sagawa}}]{14Oishi}%
  \BibitemOpen
  \bibfield  {author} {\bibinfo {author} {\bibfnamefont {T.}~\bibnamefont
  {Oishi}}, \bibinfo {author} {\bibfnamefont {K.}~\bibnamefont {Hagino}}, \
  and\ \bibinfo {author} {\bibfnamefont {H.}~\bibnamefont {Sagawa}},\ }\href
  {\doibase 10.1103/PhysRevC.90.034303} {\bibfield  {journal} {\bibinfo
  {journal} {Phys. Rev. C}\ }\textbf {\bibinfo {volume} {90}},\ \bibinfo
  {pages} {034303} (\bibinfo {year} {2014})}\BibitemShut {NoStop}%
\bibitem [{\citenamefont {Aoyama}\ \emph {et~al.}(2006)\citenamefont {Aoyama},
  \citenamefont {Myo}, \citenamefont {Kat\ifmmode~\bar{o}\else \={o}\fi{}},\
  and\ \citenamefont {Ikeda}}]{06Aoyama}%
  \BibitemOpen
  \bibfield  {author} {\bibinfo {author} {\bibfnamefont {S.}~\bibnamefont
  {Aoyama}}, \bibinfo {author} {\bibfnamefont {T.}~\bibnamefont {Myo}},
  \bibinfo {author} {\bibfnamefont {K.}~\bibnamefont {Kat\ifmmode~\bar{o}\else
  \={o}\fi{}}}, \ and\ \bibinfo {author} {\bibfnamefont {K.}~\bibnamefont
  {Ikeda}},\ }\href@noop {} {\bibfield  {journal} {\bibinfo  {journal}
  {Progress of Theoretical Physics}\ }\textbf {\bibinfo {volume} {116}},\
  \bibinfo {pages} {1} (\bibinfo {year} {2006})},\ \bibinfo {note} {and
  references therein}\BibitemShut {NoStop}%
\bibitem [{\citenamefont {Kruppa}\ \emph {et~al.}(2014)\citenamefont {Kruppa},
  \citenamefont {Papadimitriou}, \citenamefont {Nazarewicz},\ and\
  \citenamefont {Michel}}]{14Kruppa}%
  \BibitemOpen
  \bibfield  {author} {\bibinfo {author} {\bibfnamefont {A.~T.}\ \bibnamefont
  {Kruppa}}, \bibinfo {author} {\bibfnamefont {G.}~\bibnamefont
  {Papadimitriou}}, \bibinfo {author} {\bibfnamefont {W.}~\bibnamefont
  {Nazarewicz}}, \ and\ \bibinfo {author} {\bibfnamefont {N.}~\bibnamefont
  {Michel}},\ }\href {\doibase 10.1103/PhysRevC.89.014330} {\bibfield
  {journal} {\bibinfo  {journal} {Phys. Rev. C}\ }\textbf {\bibinfo {volume}
  {89}},\ \bibinfo {pages} {014330} (\bibinfo {year} {2014})}\BibitemShut
  {NoStop}%
\bibitem [{\citenamefont {Id~Betan}\ \emph {et~al.}(2002)\citenamefont
  {Id~Betan}, \citenamefont {Liotta}, \citenamefont {Sandulescu},\ and\
  \citenamefont {Vertse}}]{2002IdBetan}%
  \BibitemOpen
  \bibfield  {author} {\bibinfo {author} {\bibfnamefont {R.}~\bibnamefont
  {Id~Betan}}, \bibinfo {author} {\bibfnamefont {R.~J.}\ \bibnamefont
  {Liotta}}, \bibinfo {author} {\bibfnamefont {N.}~\bibnamefont {Sandulescu}},
  \ and\ \bibinfo {author} {\bibfnamefont {T.}~\bibnamefont {Vertse}},\ }\href
  {\doibase 10.1103/PhysRevLett.89.042501} {\bibfield  {journal} {\bibinfo
  {journal} {Phys. Rev. Lett.}\ }\textbf {\bibinfo {volume} {89}},\ \bibinfo
  {pages} {042501} (\bibinfo {year} {2002})}\BibitemShut {NoStop}%
\bibitem [{\citenamefont {Betan}\ and\ \citenamefont
  {Nazarewicz}(2012)}]{12Betan}%
  \BibitemOpen
  \bibfield  {author} {\bibinfo {author} {\bibfnamefont {R.~I.}\ \bibnamefont
  {Betan}}\ and\ \bibinfo {author} {\bibfnamefont {W.}~\bibnamefont
  {Nazarewicz}},\ }\href {\doibase 10.1103/PhysRevC.86.034338} {\bibfield
  {journal} {\bibinfo  {journal} {Phys. Rev. C}\ }\textbf {\bibinfo {volume}
  {86}},\ \bibinfo {pages} {034338} (\bibinfo {year} {2012})}\BibitemShut
  {NoStop}%
\bibitem [{\citenamefont {Id~Betan}(2017)}]{2017IdBetan}%
  \BibitemOpen
  \bibfield  {author} {\bibinfo {author} {\bibfnamefont {R.}~\bibnamefont
  {Id~Betan}},\ }\href {\doibase
  https://doi.org/10.1016/j.nuclphysa.2017.01.004} {\bibfield  {journal}
  {\bibinfo  {journal} {Nuclear Physics A}\ }\textbf {\bibinfo {volume}
  {959}},\ \bibinfo {pages} {147 } (\bibinfo {year} {2017})}\BibitemShut
  {NoStop}%
\bibitem [{\citenamefont {Myo}\ \emph {et~al.}(2008)\citenamefont {Myo},
  \citenamefont {Kikuchi}, \citenamefont {Kat\ifmmode~\bar{o}\else \={o}\fi{}},
  \citenamefont {Hiroshi},\ and\ \citenamefont {Ikeda}}]{08Myo}%
  \BibitemOpen
  \bibfield  {author} {\bibinfo {author} {\bibfnamefont {T.}~\bibnamefont
  {Myo}}, \bibinfo {author} {\bibfnamefont {Y.}~\bibnamefont {Kikuchi}},
  \bibinfo {author} {\bibfnamefont {K.}~\bibnamefont {Kat\ifmmode~\bar{o}\else
  \={o}\fi{}}}, \bibinfo {author} {\bibfnamefont {T.}~\bibnamefont {Hiroshi}},
  \ and\ \bibinfo {author} {\bibfnamefont {K.}~\bibnamefont {Ikeda}},\
  }\href@noop {} {\bibfield  {journal} {\bibinfo  {journal} {Progress of
  Theoretical Physics}\ }\textbf {\bibinfo {volume} {119}},\ \bibinfo {pages}
  {561} (\bibinfo {year} {2008})}\BibitemShut {NoStop}%
\bibitem [{\citenamefont {Bohm}\ \emph {et~al.}(1989)\citenamefont {Bohm},
  \citenamefont {Gadella},\ and\ \citenamefont {Mainland}}]{89Bohm}%
  \BibitemOpen
  \bibfield  {author} {\bibinfo {author} {\bibfnamefont {A.}~\bibnamefont
  {Bohm}}, \bibinfo {author} {\bibfnamefont {M.}~\bibnamefont {Gadella}}, \
  and\ \bibinfo {author} {\bibfnamefont {G.~B.}\ \bibnamefont {Mainland}},\
  }\href {\doibase 10.1119/1.15797} {\bibfield  {journal} {\bibinfo  {journal}
  {American Journal of Physics}\ }\textbf {\bibinfo {volume} {57}},\ \bibinfo
  {pages} {1103} (\bibinfo {year} {1989})}\BibitemShut {NoStop}%
\bibitem [{\citenamefont {Rotureau}\ \emph {et~al.}(2005)\citenamefont
  {Rotureau}, \citenamefont {Oko\l{}owicz},\ and\ \citenamefont
  {P\l{}oszajczak}}]{05Rotu}%
  \BibitemOpen
  \bibfield  {author} {\bibinfo {author} {\bibfnamefont {J.}~\bibnamefont
  {Rotureau}}, \bibinfo {author} {\bibfnamefont {J.}~\bibnamefont
  {Oko\l{}owicz}}, \ and\ \bibinfo {author} {\bibfnamefont {M.}~\bibnamefont
  {P\l{}oszajczak}},\ }\href {\doibase 10.1103/PhysRevLett.95.042503}
  {\bibfield  {journal} {\bibinfo  {journal} {Phys. Rev. Lett.}\ }\textbf
  {\bibinfo {volume} {95}},\ \bibinfo {pages} {042503} (\bibinfo {year}
  {2005})}\BibitemShut {NoStop}%
\bibitem [{\citenamefont {Rotureau}\ \emph {et~al.}(2006)\citenamefont
  {Rotureau}, \citenamefont {Oko\l{}owicz},\ and\ \citenamefont
  {P\l{}oszajczak}}]{06Rotu}%
  \BibitemOpen
  \bibfield  {author} {\bibinfo {author} {\bibfnamefont {J.}~\bibnamefont
  {Rotureau}}, \bibinfo {author} {\bibfnamefont {J.}~\bibnamefont
  {Oko\l{}owicz}}, \ and\ \bibinfo {author} {\bibfnamefont {M.}~\bibnamefont
  {P\l{}oszajczak}},\ }\href {\doibase
  http://dx.doi.org/10.1016/j.nuclphysa.2005.12.005} {\bibfield  {journal}
  {\bibinfo  {journal} {Nuclear Physics A}\ }\textbf {\bibinfo {volume}
  {767}},\ \bibinfo {pages} {13 } (\bibinfo {year} {2006})}\BibitemShut
  {NoStop}%
\bibitem [{\citenamefont {Hagen}\ \emph {et~al.}(2006)\citenamefont {Hagen},
  \citenamefont {Hjorth-Jensen},\ and\ \citenamefont {Michel}}]{06Hagen}%
  \BibitemOpen
  \bibfield  {author} {\bibinfo {author} {\bibfnamefont {G.}~\bibnamefont
  {Hagen}}, \bibinfo {author} {\bibfnamefont {M.}~\bibnamefont
  {Hjorth-Jensen}}, \ and\ \bibinfo {author} {\bibfnamefont {N.}~\bibnamefont
  {Michel}},\ }\href {\doibase 10.1103/PhysRevC.73.064307} {\bibfield
  {journal} {\bibinfo  {journal} {Phys. Rev. C}\ }\textbf {\bibinfo {volume}
  {73}},\ \bibinfo {pages} {064307} (\bibinfo {year} {2006})}\BibitemShut
  {NoStop}%
\bibitem [{\citenamefont {Kaneko}\ \emph {et~al.}(1991)\citenamefont {Kaneko},
  \citenamefont {LeMere},\ and\ \citenamefont {Tang}}]{1991Kaneko}%
  \BibitemOpen
  \bibfield  {author} {\bibinfo {author} {\bibfnamefont {T.}~\bibnamefont
  {Kaneko}}, \bibinfo {author} {\bibfnamefont {M.}~\bibnamefont {LeMere}}, \
  and\ \bibinfo {author} {\bibfnamefont {Y.~C.}\ \bibnamefont {Tang}},\ }\href
  {\doibase 10.1103/PhysRevC.44.1588} {\bibfield  {journal} {\bibinfo
  {journal} {Phys. Rev. C}\ }\textbf {\bibinfo {volume} {44}},\ \bibinfo
  {pages} {1588} (\bibinfo {year} {1991})}\BibitemShut {NoStop}%
\bibitem [{\citenamefont {Gurvitz}\ and\ \citenamefont
  {Kalbermann}(1987)}]{87Gur}%
  \BibitemOpen
  \bibfield  {author} {\bibinfo {author} {\bibfnamefont {S.~A.}\ \bibnamefont
  {Gurvitz}}\ and\ \bibinfo {author} {\bibfnamefont {G.}~\bibnamefont
  {Kalbermann}},\ }\href {\doibase 10.1103/PhysRevLett.59.262} {\bibfield
  {journal} {\bibinfo  {journal} {Phys. Rev. Lett.}\ }\textbf {\bibinfo
  {volume} {59}},\ \bibinfo {pages} {262} (\bibinfo {year} {1987})}\BibitemShut
  {NoStop}%
\bibitem [{\citenamefont {Talou}\ \emph {et~al.}(2000)\citenamefont {Talou},
  \citenamefont {Carjan}, \citenamefont {Negrevergne},\ and\ \citenamefont
  {Strottman}}]{00Talou}%
  \BibitemOpen
  \bibfield  {author} {\bibinfo {author} {\bibfnamefont {P.}~\bibnamefont
  {Talou}}, \bibinfo {author} {\bibfnamefont {N.}~\bibnamefont {Carjan}},
  \bibinfo {author} {\bibfnamefont {C.}~\bibnamefont {Negrevergne}}, \ and\
  \bibinfo {author} {\bibfnamefont {D.}~\bibnamefont {Strottman}},\ }\href
  {\doibase 10.1103/PhysRevC.62.014609} {\bibfield  {journal} {\bibinfo
  {journal} {Phys. Rev. C}\ }\textbf {\bibinfo {volume} {62}},\ \bibinfo
  {pages} {014609} (\bibinfo {year} {2000})}\BibitemShut {NoStop}%
\bibitem [{\citenamefont {Maruyama}\ \emph {et~al.}(2012)\citenamefont
  {Maruyama}, \citenamefont {Oishi}, \citenamefont {Hagino},\ and\
  \citenamefont {Sagawa}}]{12Maru}%
  \BibitemOpen
  \bibfield  {author} {\bibinfo {author} {\bibfnamefont {T.}~\bibnamefont
  {Maruyama}}, \bibinfo {author} {\bibfnamefont {T.}~\bibnamefont {Oishi}},
  \bibinfo {author} {\bibfnamefont {K.}~\bibnamefont {Hagino}}, \ and\ \bibinfo
  {author} {\bibfnamefont {H.}~\bibnamefont {Sagawa}},\ }\href {\doibase
  10.1103/PhysRevC.86.044301} {\bibfield  {journal} {\bibinfo  {journal} {Phys.
  Rev. C}\ }\textbf {\bibinfo {volume} {86}},\ \bibinfo {pages} {044301}
  (\bibinfo {year} {2012})}\BibitemShut {NoStop}%
\bibitem [{\citenamefont {Scamps}\ and\ \citenamefont
  {Hagino}(2015)}]{2015Scamp}%
  \BibitemOpen
  \bibfield  {author} {\bibinfo {author} {\bibfnamefont {G.}~\bibnamefont
  {Scamps}}\ and\ \bibinfo {author} {\bibfnamefont {K.}~\bibnamefont
  {Hagino}},\ }\href {\doibase 10.1103/PhysRevC.91.044606} {\bibfield
  {journal} {\bibinfo  {journal} {Phys. Rev. C}\ }\textbf {\bibinfo {volume}
  {91}},\ \bibinfo {pages} {044606} (\bibinfo {year} {2015})}\BibitemShut
  {NoStop}%
\bibitem [{\citenamefont {Bertulani}\ \emph {et~al.}(2007)\citenamefont
  {Bertulani}, \citenamefont {Flambaum},\ and\ \citenamefont
  {Zelevinsky}}]{07Bertulani_34}%
  \BibitemOpen
  \bibfield  {author} {\bibinfo {author} {\bibfnamefont {C.~A.}\ \bibnamefont
  {Bertulani}}, \bibinfo {author} {\bibfnamefont {V.~V.}\ \bibnamefont
  {Flambaum}}, \ and\ \bibinfo {author} {\bibfnamefont {V.~G.}\ \bibnamefont
  {Zelevinsky}},\ }\href {http://stacks.iop.org/0954-3899/34/i=11/a=006}
  {\bibfield  {journal} {\bibinfo  {journal} {Journal of Physics G: Nuclear and
  Particle Physics}\ }\textbf {\bibinfo {volume} {34}},\ \bibinfo {pages}
  {2289} (\bibinfo {year} {2007})}\BibitemShut {NoStop}%
\bibitem [{\citenamefont {Shotter}\ and\ \citenamefont
  {Shotter}(2011)}]{11Shot}%
  \BibitemOpen
  \bibfield  {author} {\bibinfo {author} {\bibfnamefont {A.~C.}\ \bibnamefont
  {Shotter}}\ and\ \bibinfo {author} {\bibfnamefont {M.~D.}\ \bibnamefont
  {Shotter}},\ }\href {\doibase 10.1103/PhysRevC.83.054621} {\bibfield
  {journal} {\bibinfo  {journal} {Phys. Rev. C}\ }\textbf {\bibinfo {volume}
  {83}},\ \bibinfo {pages} {054621} (\bibinfo {year} {2011})}\BibitemShut
  {NoStop}%
\end{thebibliography}

\end{document}